\documentclass[twocolumn,preprintnumbers,amsmath,amssymb]{revtex4}
\usepackage{amsfonts}
\usepackage{amssymb,epic,eepic}
\usepackage{amsmath}
\usepackage{dcolumn}
\usepackage{bm}
\newcommand{\hr}{{\mathcal H}}
\newcommand{\cn}{{\mathcal N }}

\newcommand{\cm}{{\mathcal M}}
\newcommand{\cs}{{\mathcal S}}
\newcommand{\crr}{{\mathcal R}}
\newcommand{\fr}{{\mathcal F}}
\newcommand{\gr}{{\mathcal G}}
\newcommand{\fri}{{\mathfrak I}}
\newcommand{\kr}{{\mathcal K}}

\newcommand{\cc}{{\mathbb C}}
\newcommand{\rr}{{\mathbb R}}
\newcommand{\nn}{{\mathbb N}}
\newcommand{\idn}{\mathbf{1}}

\newcommand{\eps}{{\varepsilon}}        
\newcommand{\vphi}{{\varphi}}           

\newtheorem{theorem}{Theorem}[section]         
\newtheorem{lemma}[theorem]{Lemma}             
\newtheorem{corollary}[theorem]{Corollary}     
\date{August 7, 2008}
\begin{document}

\title{On Quantum Capacity of Compound Channels}

\author{Igor Bjelakovi\'c}
\email{igor.bjelakovic@mk.tu-berlin.de}
\affiliation{Heinrich-Hertz-Chair for Mobile Communications, Technische Universit\"at Berlin}
\altaffiliation[The first two authors are also with ]{Institut f\"ur Mathematik, Technische Universit\"at Berlin}

\author{Holger Boche}
\email{holger.boche@mk.tu-berlin.de}
\affiliation{Heinrich-Hertz-Chair for Mobile Communications, Technische Universit\"at Berlin}
\altaffiliation{Institut f\"ur Mathematik Technische Universit\"at Berlin}

\author{Janis N\"otzel}
\email{janis.noetzel@mk.tu-berlin.de}
\affiliation{Heinrich-Hertz-Chair for Mobile Communications, Technische Universit\"at Berlin}

\&\\

\begin{abstract}
In this paper we address the issue of universal or robust communication over quantum channels. Specifically, we consider memoryless communication scenario with channel uncertainty which is an analog of compound channel in classical information theory.
We determine the quantum capacity of finite compound channels and arbitrary compound channels with informed decoder. Our approach in the finite case is based on the observation that perfect channel knowledge at the decoder does not increase the capacity of finite quantum compound channels. As a consequence we obtain coding theorem for finite quantum averaged channels, the simplest class of channels with long-term memory. The extension of these results to quantum compound channels with uninformed encoder and decoder, and infinitely many constituents remains an open problem.
\end{abstract}

\maketitle
\section{\label{sec:introduction}Introduction}
The determination of capacities of quantum channels in various settings is one of the most fundamental tasks of quantum information theory. The classical capacity of memoryless quantum channels has been identified in seminal work by Holevo \cite{holevo} and Schumacher and Westmoreland \cite{schumacher-westmoreland}. A much sharper version of that coding theorem was obtained subsequently and independently by Winter \cite{winter} who proved the strong converse and achievability with von Neumann measurements, and Ogawa and Nagaoka \cite{ogawa-nagaoka} who gave another proof of the strong converse to the coding theorem.
Concerning quantum capacity, Lloyd \cite{lloyd} provided a strong heuristic evidence that the quantum capacity of quantum memoryless channels is given by the regularized coherent information. The follow-up work of Shor \cite{shor} and Devetak \cite{devetak} gradually removed the residual ambiguity and culminated in a rigorous proof of the direct part of the coding theorem for memoryless quantum channels in \cite{devetak}. The converse coding theorem was already done by Barnum, Knill, and Nielsen \cite{bkn}. \\
The underlying assumptions on the quantum communication model considered in that fundamental work are a) that the quantum channel is perfectly known, and b) that there are no correlations between successive channel uses, i.e. the channel is memoryless. Both postulates are rarely fulfilled for real world communication systems. E.g. it is practically impossible to determine all parameters the channel depends on with infinite precision. Thus, we are forced to design protocols which work well for communication scenarios with channel uncertainty. Similarly, it is definitely too optimistic to presuppose that the channel under consideration operates in a memoryless fashion.\\
In this work we relax the first assumption while sustaining the memoryless character of the communication. Our model can be described as follows: We are given an \emph{arbitrary} set $\fri$ of memoryless quantum channels. Both sender and receiver merely know that the actual channel belongs to this set. Consequently, they are forced to find and use codes that are reliable for the whole set $\fri$ of channels. This is exactly the quantum analog of compound channels from classical information theory. Moreover this is a channel version of the universal quantum data compression which has been introduced by Jozsa, M., P., and R. Horodecki \cite{jozsa-horodecki}.\\
The main results of the paper are:
\begin{enumerate}
\item The determination of the quantum capacity of compound channels in the case $|\fri|<\infty$. Here we first prove the coding theorem for finite $\fri$ assuming that the decoder knows which channel is in use (informed decoder for short). Then we show that each rate achievable in this situation is also achievable without such prior knowledge. As a consequence we can determine the quantum capacity of finite averaged channels. Eventually, we determine the capacity of quantum compound channel with informed encoder in the case $|\fri|<\infty$. Denoting the quantum capacities of $\fri$ with informed encoder by $Q_{IE}(\fri)$, with informed decoder by $Q_{ID}(\fri)$, and without any prior information by $Q(\fri)$, our results in this part of the paper can be roughly summarized by 
\[Q(\fri)=Q_{ID}(\fri)\le Q_{IE}(\fri).  \]
It can be shown, by generalizing an example given by Blackwell, Breiman, and Thomasian \cite{blackwell}, that the inequality is strict in general.
\item We prove the coding theorem for arbitrary $\fri$ with informed decoder. Unfortunately, the techniques we employ do not allow an extension of the result for finite sets of channels to arbitrary $\fri$ without channel knowledge at the decoder.
\end{enumerate}
\subsection{Related Work}
The capacity of compound channels in the classical setting was determined by Blackwell, Breiman, and Thomasian \cite{blackwell} and Wolfowitz \cite{wolfowitz-compound, wolfowitz-book}. It was Wolfowitz \cite{wolfowitz-compound, wolfowitz-book} who realized the importance of channel knowledge at the encoder and the decoder of the compound channel and its theoretical implications. He showed, that the assumption of informed decoder does not imply a higher capacity of the compound channel. His approach was to establish first the direct part for the compound channel without channel knowledge at the decoder and to show that the upper bound of the converse part with informed decoder coincides with the lower bound obtained in the first step. So, our proof strategy described above is exactly  the reverse to that of Wolfowitz.\\
In a beautiful paper \cite{datta-dorlas} Datta and Dorlas determined the classical capacity of \emph{finite} averaged quantum channels, a channel model which is equivalent to \emph{finite} quantum compound channels, i.e. $|\fri|<\infty$. Somewhat later an independent approach in the finite case was given in \cite{farkas} by Farkas.
 Subsequently, two of us \cite{bb-compound} were able to identify the classical capacity of \emph{arbitrary} quantum compound channels and, as a consequence thereof, the classical capacity of arbitrary averaged channels. Recently Hayashi \cite{hayashi-universal} obtained a similar result with a completely different proof technique. He used the Schur-Weyl duality and the packing lemma from \cite{csiszar}.
\subsection{Outline}
The paper is organized as follows: Section \ref{sec:codes-and-capacity} provides the necessary definitions concerning compound channels and codes in scenarios with and without informed decoder. The following section \ref{sec:one-shot-section} contains a modified version of Klesse's \cite{klesse} one-shot coding result which applies to finite arithmetic averages of quantum channels. Since good codes for such averaged channels are also good for quantum channels that are averaged over we obtain the basic building block of the coding theorem for finite compound channels with informed decoder which is derived in section \ref{sec:finite-direct-part}. Moreover, using the elegant channel estimation strategy developed by Datta and Dorlas \cite{datta-dorlas} we are able to convert the codes for finite quantum compound channels with informed decoder into codes that do not require such prior information at the decoder. Section \ref{sec:finite-direct-part} concludes with the coding theorem for finite quantum compound channels with informed encoder.\\
Discretization arguments via $\tau$-nets and some simple approximation results concerning the coherent information and entanglement fidelity are presented in section \ref{sec:nets}. They are the basis for section \ref{sec:direct-part} in which we carry out the extension of the capacity result with informed decoder to the case of arbitrary set $\fri$.
\subsection{Notation and Conventions}
All Hilbert spaces are assumed to have finite dimension and are over the field $\cc$. $\mathcal{S}(\hr)$ is the set of states, i.e. positive semi-definite operators with trace $1$ acting on the Hilbert space $\hr$.
The set of completely positive trace preserving (CPTP) maps between the operator spaces $\mathcal{B}(\hr)$ and $\mathcal{B}(\kr)$ is denoted by $\mathcal{C}(\hr,\kr)$. $\mathcal{C}^{\downarrow}(\hr,\kr)$ stands for the set of completely positive trace decreasing maps between $\mathcal{B}(\hr)$ and $\mathcal{B}(\kr)$. Throughout the paper we use base two logarithms which are denoted by $\log$. The von Neumann entropy of a state $\rho\in\mathcal{S}(\hr)$ is given by
\[S(\rho):=-\textrm{tr}(\rho \log\rho).  \]
The coherent information for $\cn\in \mathcal{C}(\hr,\kr) $ and $\rho\in\mathcal{S}(\hr)$ is defined by
\[I_c(\rho, \cn):=S(\cn (\rho))- S( (id_{\hr}\otimes \cn)(|\psi\rangle\langle \psi|)  ),  \]
where $\psi\in\hr\otimes \hr$ is a purification of the state $\rho$.\\
A useful equivalent definition of $I_c(\rho,\cn)$ is given in terms of $\cn$ and the complementary channel $\mathcal{E}$: Due to Steinspring's dilation theorem $\cn$ can be represented as $\cn(\rho)=\textrm{tr}_{\hr_e}(v\rho v^{\ast})$ for $\rho \in\mathcal{S}(\hr) $ where $v:\hr \to\kr \otimes \hr_e$ is a linear isometry. The complementary channel $\mathcal{E}\in\mathcal{C}(\hr,\hr_e)$ to $\cn$ is given by
\[\mathcal{E}(\rho):= \textrm{tr}_{\hr}(v\rho v^{\ast})\qquad (\rho \in\mathcal{S}(\hr)).  \]
The coherent information can then be written as
\[I_c(\rho, \cn)=S(\cn (\rho))- S(\mathcal{E}(\rho)). \]
As a last notational issue we recall the definition of entanglement fidelity. For $\rho\in\mathcal{S}(\hr)$ and $\cn\in \mathcal{C}^{\downarrow}(\hr,\kr)$ it is given by
\[F_e(\rho,\cn):=\langle\psi, (id_{\hr}\otimes \cn)(|\psi\rangle\langle \psi|)     \psi\rangle,  \]
with a purification $\psi\in\hr\otimes \hr$ of the state $\rho$.
\section{\label{sec:codes-and-capacity} Codes and Capacity}
Let $\fri\subset \mathcal{C}(\hr,\kr)$. The memoryless compound channel associated with $\fri$ is given by the family $\{\cn^{\otimes l}:\mathcal{S}(\hr^{\otimes l})\to\mathcal{S}(\kr^{\otimes l})  \}_{\l\in\nn,\cn\in\fri}$. In the rest of the paper we will write simply $\fri$ for that family.\\
Most of the time we will deal with compound channels with informed decoder, i.e. we suppose that the receiver knows which channel is in use. The definition of codes in this situation is as follows:\\
 An $(l,k_l)-$ \emph{code} for $\fri$ with \emph{informed decoder} consists of a subspace $\fr_l\subset\hr^{\otimes l}$ with $k_l=\dim \fr_l$ and a family of CPTP maps $\{ \crr_{\cn}^l:\mathcal{S}(\kr^{\otimes l})\to\mathcal{S}(\hr^{\otimes l}) \}_{\cn\in\fri}$.\\
A nonnegative number $R$ is called an achievable rate for $\fri$ with informed decoder if there is a sequence of $(l,k_l)$-codes such that
\begin{enumerate}
\item $\liminf_{l\to\infty}\frac{1}{l}\log k_l\ge R$, and
\item $\lim_{l\to\infty}\inf_{\cn\in\fri}F_e(\pi_{\fr_l},\crr_{\cn}^l\circ \cn^{\otimes l})=1 $
\end{enumerate}
hold with $\pi_{\fr_l}:=\frac{p_{\fr_l}}{\textrm{tr}(p_{\fr_l})}$ where $p_{\fr_l}$ is the orthogonal projection onto $\fr_l$.\\
The \emph{capacity} $Q_{ID}(\fri)$ of the memoryless compound channel $\fri$ with informed decoder is given by
\begin{eqnarray*}
 Q_{ID}(\fri):=\sup\{R\in\rr_{+}&:& R \textrm{ is achievable for } \fri\\
&&  \textrm{with informed decoder}   \}.
\end{eqnarray*}
The definition of codes in the situation of an informed encoder is straightforward:\\
We endow the encoder with a set of subspaces, one for each channel in $\fri$.\\
An $(l,k_l)-$ \emph{code} for $\fri$ with \emph{informed encoder} thus consists of subspaces $\{\fr_l^{\cn}\}_{\cn\in\fri}$ with $\fr_l^\cn\subset\hr^{\otimes l}$ and $k_l=\dim \fr_l^\cn$ for all $\cn\in\fri$ and a recovery operation $\crr^l:\mathcal{S}(\kr^{\otimes l})\to\mathcal{S}(\hr^{\otimes l})$.\\
A nonnegative number $R$ is called an achievable rate for $\fri$ with informed encoder if there is a sequence of $(l,k_l)$-codes such that
\begin{enumerate}
\item $\liminf_{l\to\infty}\frac{1}{l}\log k_l\ge R$, and
\item $\lim_{l\to\infty}\inf_{\cn\in\fri}F_e(\pi_{\fr_l^\cn},\crr^l\circ \cn^{\otimes l})=1 $
\end{enumerate}
hold with $\pi_{\fr_l^\cn}:=\frac{p_{\fr_l^\cn}}{\textrm{tr}(p_{\fr_l^\cn})}$ where $p_{\fr_l^\cn}$ is the orthogonal projection onto $\fr_l^\cn$.\\
The \emph{capacity} $Q_{IE}(\fri)$ of the memoryless compound channel $\fri$ with informed encoder is given by
\begin{eqnarray*}
 Q_{IE}(\fri):=\sup\{R\in\rr_{+}&:& R \textrm{ is achievable for } \fri\\
&&  \textrm{with informed encoder}   \}.
\end{eqnarray*}
Codes and capacity for the memoryless compound channel $\fri$ with \emph{uninformed encoder and decoder} are defined in a similar fashion. The only change is that we do not allow the recovery operation to depend on $\cn\in\fri$. I.e.  An $(l,k_l)-$ \emph{code} for $\fri$ is a pair $(\fr_l, \crr^l)$ where $\fr_l$ is a subspace of $\hr^{\otimes l}$ with $k_l=\dim \fr_l$ and $\crr^l\in\mathcal{C}(\kr^{\otimes l},\hr^{\otimes l})$.\\
A nonnegative number $R$ is called an achievable rate for $\fri$ if there is a sequence of $(l,k_l)$-codes such that
\begin{enumerate}
\item $\liminf_{l\to\infty}\frac{1}{l}\log k_l\ge R$, and
\item $\lim_{l\to\infty}\inf_{\cn\in\fri}F_e(\pi_{\fr_l},\crr^l\circ \cn^{\otimes l})=1 $.
\end{enumerate}
The \emph{capacity} $Q(\fri)$ of the memoryless compound channel $\fri$ is given by
\begin{equation*}
 Q(\fri):=\sup\{R\in\rr_{+}: R \textrm{ is achievable for } \fri   \}.
\end{equation*}
It is clear from the definitions given above that
\begin{equation}\label{q-le-qid}
 Q(\fri)\le \min\{ Q_{IE}(\fri), Q_{ID}(\fri)
\end{equation}
holds for each $\fri\subset\mathcal{C}(\hr,\kr)$.\\
Throughout the paper we will use the close relationship between the quantum compound channels with $|\fri|<\infty$ and (finite) averaged channels. The latter are described as follows: Given any finite set $\fri=\{ \cn_1,\ldots \cn_N \}\subset\mathcal{C}(\hr,\kr)$, real numbers $\lambda_1,\ldots,\lambda_N> 0$ with $\sum_{i=1}^N\lambda_i=1$, and $l\in\nn$ we set
\[\cn^l:=\sum_{i=1}^N \lambda_i\cn_i^{\otimes l}.  \]
The averaged channel $\mathfrak{A}_{\fri,\lambda}$ associated with $\fri$ and $\lambda:=(\lambda_1,\ldots,\lambda_N)$ is then the family $\{\cn^l:\mathcal{S}(\hr^{\otimes l})\to\mathcal{S}(\kr^{\otimes l})  \}_{l\in\nn}$. The codes and capacities, $Q_{ID}(\mathfrak{A}_{\fri,\lambda} )$ and $Q(\mathfrak{A}_{\fri,\lambda} ) $, with or without informed decoder are defined in a similar fashion as for compound channels.\\
Clearly, an $(l,k_l)$-code for $\fri$ with informed decoder and
\[F_e(\pi_{\fr_l}, \crr_i^l\circ \cn_i^{\otimes l})\ge 1-\eps \quad \forall i\in\{1,\ldots, N  \} \]
implies
\begin{equation}\label{eq:basic-relation}
F_e(\pi_{\fr_l}, \sum_{i=1}^n\lambda_i \crr_i^l\circ \cn_i^{\otimes l})\ge 1-\eps.
\end{equation}
Conversely, (\ref{eq:basic-relation}) implies
\[F_e(\pi_{\fr_l}, \crr_i^l\circ \cn_i^{\otimes l})\ge 1-\frac{\eps}{\lambda_i} \quad \forall i\in\{1,\ldots, N  \} .  \]
Similar relation holds in the case of uninformed decoder. Thus we are led to the following simple consequence which will be used freely in the rest of the paper.
\begin{lemma}\label{finite-averaged-vs-compound}
Let $\fri=\{ \cn_1,\ldots \cn_N \}\subset\mathcal{C}(\hr,\kr)$ and real numbers $\lambda_1,\ldots,\lambda_N> 0$ with $\sum_{i=1}^N\lambda_i=1$ be given. Then
\[Q_{ID}(\fri)=Q_{ID}(\mathfrak{A}_{\fri,\lambda} )\quad \textrm{and}\quad  Q(\fri)=Q(\mathfrak{A}_{\fri,\lambda} ) .  \]
\end{lemma}

\section{\label{sec:one-shot-section}One-Shot Coding Result}
Our coding result will be based on a one-shot result which is an extension of that for a single channel obtained in \cite{klesse} by Klesse. A similar approach for a single channel is given in the paper \cite{hayden} by Hayden, Horodecki, Yard, and Winter.\\
We start with subspaces $\fr\subset \gr \subset \hr$ of a given Hilbert space $\hr$ and states $\pi_{\fr}:=p/\textrm{tr}(p)$  and $\pi_{\gr}:=q/\textrm{tr}(q)$ where $p$ and $q$ are the projections onto $\fr$ and $\gr$ respectively. Recall the definition of the code entanglement fidelity as
\[F_{c,e}(\pi_{\fr},\cn):=\max_{\crr \in \mathcal{C}(\kr,\hr)}F_e(\pi_{\fr},\crr\circ\cn), \]
where we admit any completely positive trace decreasing (CPTD) map $\cn:\mathcal{B}(\hr)\to \mathcal{B}(\kr)$. The set of completely positive trace decreasing maps is denoted by $\mathcal{C}^{\downarrow}(\hr,\kr)$. Klesse's main result in \cite{klesse} can be stated as follows:
\begin{theorem}[Klesse \cite{klesse}]\label{klesse-theorem}
Let the Hilbert space $\hr$ be given and consider subspaces $\fr\subset\gr\subset \hr$ with $\dim \fr=k$. Then for any $\cn\in \mathcal{C}^{\downarrow}(\hr,\kr)$ allowing a representation with $n$ Kraus operators we have
\[\int_{\mathfrak{U}(\gr)}F_{c,e}(u\pi_{\fr}u^{\ast},\cn)du\ge \textrm{tr}(\cn(\pi_{\gr}))-\sqrt{k\cdot n}||\cn(\pi_{\gr})||_2,  \]
where $\mathfrak{U}(\gr) $ denotes the group of unitaries acting on $\gr$ and $du$ indicates that the integration is with respect to the Haar measure on $\mathfrak{U}(\gr) $.
\end{theorem}
Our main goal in this section will be a variant of Theorem \ref{klesse-theorem} which applies to finite arithmetic averages of CPTD maps.\\
Let $\cn_1,\ldots ,\cn_N\in \mathcal{C}^{\downarrow}(\hr,\kr) $ and set
\[\cn:=\frac{1}{N}\sum_{i=1}^{N}\cn_i\in \mathcal{C}^{\downarrow}(\hr,\kr).  \]
As in Theorem \ref{klesse-theorem} we consider subspaces $\fr\subset\gr\subset \hr$ and states $\pi_{\fr}$ and $\pi_{\gr}$.\\
Let $\psi\in\hr_a\otimes \hr$, $\hr_a=\hr$, be a purification of $\pi_{\fr}$ and consider a Stinespring dilation of the channel $\cn$ given by
\begin{equation}\label{stinespring-1}
  \cn (\ \cdot\ )=\textrm{tr}_{\hr_e}((\idn_{\hr}\otimes p_e)v(\ \cdot\ )v^{\ast} ),
\end{equation}
where $\hr_e$ is a suitable finite-dimensional Hilbert space, $p_e$ is a projection onto a subspace of $\hr_e$, and $v:\hr \to\kr\otimes \hr_e $ is an isometry. \\
Let us define a pure state on $\hr_a\otimes \kr\otimes \hr_e$ by the formula
\[  \psi':= \frac{1}{\sqrt{\textrm{tr}(\cn(\pi_{\fr}))}}(\idn_{\hr_a\otimes \hr}\otimes p_a)(\idn_{\hr_a}\otimes v) \psi.  \]
We set
\[\rho':=\textrm{tr}_{\hr_a\otimes \hr_e}(|\psi'\rangle\langle \psi'|),\quad \rho'_{ae}:= \textrm{tr}_{\kr}(|\psi'\rangle\langle \psi'|), \]
and
\[\rho_a:=\textrm{tr}_{\hr}(|\psi\rangle\langle \psi|),\quad \rho'_e:=\textrm{tr}_{\hr_a\otimes \kr}(|\psi'\rangle\langle \psi'|).  \]
We borrow the following lemma from \cite{klesse, hayden} which will be the basis of the result we are going to derive in this section.
\begin{lemma}[Cf. \cite{klesse, hayden}]\label{decoupling-lemma}
For any $\cn\in \mathcal{C}^{\downarrow}(\hr,\kr) $ there exists a recovery operation $\crr \in \mathcal{C}(\kr,\hr) $ with
\[F_e(\rho, \crr\circ \cn)\ge w-||w\rho'_{ae}-w\rho_a\otimes \rho'_e||_1,  \]
where $w=\textrm{tr}(\cn(\pi_{\fr}))$.
\end{lemma}
The next, simple lemma is needed in the proof of theorem \ref{convex-klesse}.
\begin{lemma}\label{matrix-lemma}
Let $L$ and $D$ be $N\times N$ matrices with non-negative entries which satisfy
\begin{equation}\label{matrix-1}
 L_{jl}\le L_{jj}, \quad L_{j l}\le L_{ll},
\end{equation}
and
\begin{equation}\label{matrix-2}
 D_{jl}\le \max \{ D_{jj}, D_{ll} \}
\end{equation}
for all $j,l\in \{ 1,\ldots, N \}$. Then
\[\sum_{j,l=1}^{N}\frac{1}{N}\sqrt{L_{jl}D_{jl}}\le 2\sum_{j=1}^{N}\sqrt{L_{jj}D_{jj}}.  \]
\end{lemma}
\emph{Proof.} Note that (\ref{matrix-2}) implies
\begin{equation}\label{matrix-3}
  D_{jl}\le D_{jj}+ D_{ll}.
\end{equation}
Therewith we obtain
\begin{eqnarray}
  \sum_{j,l=1}^{N}\frac{1}{N}\sqrt{L_{jl}D_{jl}}&\le &\sum_{j,l=1}^{N}\frac{1}{N}\sqrt{L_{jl}(D_{jj}+ D_{ll})  }\label{matrix-4}\\
&\le & \sum_{j,l=1}^{N}\frac{1}{N}\sqrt{L_{jj}D_{jj}+L_{ll} D_{ll}  }\label{matrix-5}\\
&\le & \sum_{j,l=1}^{N}\frac{1}{N}\left(\sqrt{L_{jj}D_{jj}}+\sqrt{L_{ll} D_{ll}  }\right)\label{matrix-6}\\
&=& 2\sum_{j=1}^{N}\sqrt{L_{jj}D_{jj}},\nonumber
\end{eqnarray}
where in (\ref{matrix-4}) we have used (\ref{matrix-3}), in (\ref{matrix-5}) we employed (\ref{matrix-1}), and (\ref{matrix-6}) holds because $\sqrt{a+b}\le \sqrt{a}+\sqrt{b}$ for all non-negative real numbers $a,b$.
\begin{flushright}$\Box$\end{flushright}
Our main result in this section can be stated as follows:
\begin{theorem}\label{convex-klesse}
Let the Hilbert space $\hr$ be given and consider subspaces $\fr\subset\gr\subset \hr$ with $\dim \fr=k$. For any choice of $\cn_1,\ldots \cn_N\in \mathcal{C}^{\downarrow}(\hr,\kr) $ each allowing a representation with $n_j$ Kraus operators, $j=1,\ldots , N$, we set
\[\cn:=\frac{1}{N}\sum_{j=1}^{N}\cn_j.  \]
Then
\begin{eqnarray*}
 \int_{\mathfrak{U}(\gr)}F_{c,e}(u\pi_{\fr}u^{\ast},\cn)du&\ge& \textrm{tr}(\cn(\pi_{\gr})) \\
&&- 2 \sum_{j=1}^N \sqrt{k n_j }||\cn_j (\pi_{\gr})||_2.
\end{eqnarray*}
\end{theorem}
\emph{Proof.} We can assume without loss of generality that the numbering of the channels is chosen in such a way that $n_1\le n_2\le \ldots \le n_N$ holds for the numbers of Kraus operators of the maps $\cn_1,\ldots, \cn_N$. From lemma \ref{decoupling-lemma} we know that there is a recovery operation $\crr$ such that
\begin{equation}\label{eq:convex-klesse-1}
 F_e(\pi_{\fr}, \crr\circ \cn)\ge w-||w\rho'_{ae}-w\rho_a\otimes \rho'_e||_1,
\end{equation}
where we have used the notation introduced in the paragraph preceeding lemma \ref{decoupling-lemma}.\\
For each $j\in\{1,\ldots ,N\}$ let $\{a_{j,i}\}_{i=1}^{n_j}$ be the set of Kraus operators of $\cn_j$. Let $\{f_1,\ldots, f_N  \}$ and $\{e_1,\ldots ,e_{n_N}  \}$ be arbitrary orthonormal bases of $\cc^{N}$ and $\cc^{n_N}$. Let the projection $p_e$ and the isometry $v$ in (\ref{stinespring-1}) be chosen in such a way that for each $\phi\in \hr$ the relation
\begin{equation}\label{stinespring-2}
(\idn_{\hr}\otimes p_e)v \phi=\sum_{j=1}^{N}\sum_{i=1}^{n_j}\frac{1}{\sqrt{N}}(a_{j,i}\phi)\otimes e_i\otimes f_j,
\end{equation}
holds. For a purification $\psi\in \hr_a\otimes \hr$ of the state $\pi_{\fr}$ we consider a Schmidt representation
\[\psi=\frac{1}{\sqrt{k}}\sum_{m=1}^{k}h_m\otimes g_m,  \]
with suitable orthonormal systems $\{ h_1,\ldots ,h_k \}$ and $\{ g_1,\ldots, g_k  \}$.
A calculation identical to that performed by Klesse \cite{klesse} shows that the states on the right hand side of (\ref{eq:convex-klesse-1}) can be expressed with the help of representation (\ref{stinespring-2}) as
\begin{equation}\label{eq:convex-klesse-2}
  w\rho'_{ae}=\frac{1}{k}\sum_{j,l=1}^{N}\sum_{i,r=1}^{n_j,n_l}\sum_{s,t=1}^{k}\frac{\textrm{tr}(a_{j,i}|g_s\rangle\langle g_t|a_{l,r}^{\ast} )}{N}
|x_{s,i,j}\rangle \langle x_{t,r,l}|,
\end{equation}
with $x_{s,i,j}:=h_s\otimes e_i\otimes f_j$, and
\begin{equation}\label{eq:convex-klesse-3}
  w\rho_a\otimes \rho'_e=\sum_{j,l=1}^{N}\sum_{i,r=1}^{n_j,n_l}\frac{\textrm{tr}(a_{j,i}\pi_{\fr}a_{l,r}^{\ast}  )}{kN}\rho_a\otimes |y_{i,j}\rangle \langle y_{r,l}|,
\end{equation}
where $y_{i,j}:=e_i\otimes f_j$.\\
If we perform the unitary conjugation induced by the unitary map $x_{s,i,j}=h_s\otimes e_i\otimes f_j\mapsto x'_{s,i,j}= g_s\otimes e_i\otimes f_j $ followed by the complex conjugation of the matrix elements with respect to the matrix units $\{|x'_{s,i,j}\rangle\langle x'_{t,k,l}|  \}_{s,i,j,t,k,l }$ we obtain an anti-linear isometry $I$ with respect to the metrics induced by the trace distances on the operator spaces under consideration. A calculation identical to that in \cite{klesse} shows that under this isometry the sub-normalized states in (\ref{eq:convex-klesse-2}) and (\ref{eq:convex-klesse-3}) transform to
\begin{equation}\label{eq:convex-klesse-4}
  I(w \rho'_{ae})=\frac{1}{k N}\sum_{j,l=1}^{N}\sum_{i,r=1}^{n_j,n_l}pa_{j,i}^{\ast}a_{l,r}p\otimes |y_{i,j}\rangle \langle y_{r,l}|,
\end{equation}
and
\begin{equation}\label{eq:convex-klesse-5}
  I(w\rho_a\otimes \rho'_e )=\frac{1}{k}\sum_{j,l=1}^{N}\sum_{i,r=1}^{n_j,n_l}\frac{\textrm{tr}(pa_{j,i}^{\ast}a_{l,r}p )}{kN}p\otimes |y_{i,j}\rangle \langle y_{r,l}|,
\end{equation}
with $p=k\pi_{\fr}$ and $y_{i,j}=e_i\otimes f_j $ for $j=1,\ldots, N$ and $i=1,\ldots, n_j$. In summary, using the isometry $I$, (\ref{eq:convex-klesse-4}), and (\ref{eq:convex-klesse-5}) the inequality (\ref{eq:convex-klesse-1}) can be formulated as
\begin{equation}\label{eq:convex-klesse-6}
  F_e(\pi_{\fr}, \crr\circ \cn)\ge w-||D(p)||_1,
\end{equation}
with $w=\textrm{tr}(\cn(\pi_{\fr}))$ and
\[D(p):= \sum_{j,l=1}^{N}\frac{1}{N}\sum_{i,r=1}^{n_j,n_l}D_{(ij)(rl)}(p)  \otimes |e_i\rangle \langle e_r|\otimes |f_j\rangle\langle f_l|  \]
where
\[D_{(ij)(rl)}(p):= \frac{1}{k}\left(pa_{j,i}a_{l,r}^{\ast}p- \frac{1}{k}\textrm{tr}(pa_{j,i}^{\ast}a_{l,r}p )p\right) . \]
Let us define
\begin{equation}\label{eq:convex-klesse-7}
D_{j,l}(p):= \sum_{i=1,k=1}^{n_j,n_l}D_{(ij)(kl)}(p)\otimes|e_i\rangle \langle e_k|\otimes |f_j\rangle\langle f_l|.
\end{equation}
The triangle inequality for the trace norm yields
\begin{eqnarray}\label{eq:convex-klesse-8}
  ||D(p)||_1&\le& \sum_{j,l=1}^{N}\frac{1}{N}||D_{j,l}(p)||_1 \nonumber\\
&\le & \sum_{j,l=1}^{N}\frac{1}{N}\sqrt{k\min \{n_j,n_l  \}} ||D_{j,l}(p)||_2,\nonumber\\
&=& \sum_{j,l=1}^{N}\frac{1}{N}\sqrt{k\min \{n_j,n_l  \} ||D_{j,l}(p)||_2^2},
\end{eqnarray}
where the second line is justified by the standard relation between the trace and Hilbert-Schmidt  norm, $||a ||_1\le \sqrt{d}||a ||_2$, $d$ being the number of non-zero singular values of $a$.\\
In the next step we will compute $||D_{j,l}(p)||_2^2 $. A glance at (\ref{eq:convex-klesse-7}) shows that
\begin{equation}\label{eq:convex-klesse-9}
  (D_{j,l}(p))^{\ast}=\sum_{i=1,k=1}^{n_j,n_l}(D_{(ij)(kl)}(p))^{\ast}\otimes|e_k\rangle \langle e_i|\otimes |f_l\rangle\langle f_j|,
\end{equation}
and consequently we obtain
\begin{eqnarray}\label{eq:convex-klesse-10}
||D_{j,l}(p)||_2^2&=& \textrm{tr}((D_{j,l}(p) )^{\ast}D_{j,l}(p) )  \nonumber\\
&=& \sum_{i=1,r=1}^{n_j,n_l}\textrm{tr}( (D_{(ij)(kl)}(p))^{\ast} D_{(ij)(kl)}(p))\nonumber\\
&=& \frac{1}{k^2}\sum_{i=1,r=1}^{n_j,n_l}\{\textrm{tr}(p (a_{j,i}^{\ast}a_{l,r})^{\ast}p a_{j,i}^{\ast}a_{l,r})\nonumber\\
& &-\frac{1}{k}|\textrm{tr}(pa_{j,i}^{\ast}a_{l,r} )|^2 \}.
\end{eqnarray}
Let $U$ be a random variable taking values in $\mathfrak{U}(\gr)$ according to the Haar measure of $\mathfrak{U}(\gr) $. Then we can infer from (\ref{eq:convex-klesse-8}) that
\begin{equation}\label{eq:convex-klesse-11}
  \mathbb{E}( ||D(UpU^{\ast})||_1 )\le \sum_{j,l=1}^{N}\frac{1}{N}\sqrt{L_{jl}\mathbb{E}(||D_{j,l}(UpU^{\ast}) ||_2^2) },
\end{equation}
where we have used the concavity of the function $\sqrt{\ \cdot \ }$ and Jensen's inequality, and moreover, we abbreviated $k\min\{n_j, n_l  \}$ by $L_{jl}$. Now, starting with (\ref{eq:convex-klesse-10}) and arguing as Klesse \cite{klesse} we obtain that
\begin{eqnarray}\label{eq:convex-klesse-12}
 \mathbb{E}(||D_{j,l}(UpU^{\ast}) ||_2^2) &\le& \textrm{tr}(\cn_j(\pi_{\gr})\cn_l(\pi_{\gr}) )\nonumber\\
&=& \langle\cn_j(\pi_{\gr}),\cn_l(\pi_{\gr})\rangle_{HS} ,
\end{eqnarray}
where $\langle \ \cdot \ ,\ \cdot \ \rangle_{HS} $ denotes the Hilbert-Schmidt inner product.
Similarly
\begin{equation}\label{eq:convex-klesse-13}
  \mathbb{E}(\textrm{tr}(\cn(U\pi_{\fr}U^{\ast})))=\textrm{tr}(\cn(\pi_{\gr})).
\end{equation}
Now, using (\ref{eq:convex-klesse-1}), (\ref{eq:convex-klesse-6}), (\ref{eq:convex-klesse-10}), (\ref{eq:convex-klesse-11}), (\ref{eq:convex-klesse-12}), and (\ref{eq:convex-klesse-13}) we arrive at
\begin{eqnarray}\label{eq:convex-klesse-14}
 \mathbb{E} (F_{c,e}(U\pi_{\fr}U^{\ast},\cn))&\ge& \textrm{tr}(\cn (\pi_{\gr}))\nonumber\\
&& -\sum_{j,l=1}^{N}\frac{1}{N}\sqrt{L_{jl}D_{jl}},
\end{eqnarray}
where for $j,l\in\{1,\ldots, N  \}$ we introduced the abbreviations
\[L_{jl}=k\min\{n_j, n_l  \},  \]
and
\[ D_{jl}:= \langle\cn_j(\pi_{\gr}),\cn_l(\pi_{\gr})\rangle_{HS}. \]
It is obvious that
\[  L_{jl}\le L_{jj} \quad \textrm{and}\quad  L_{j l}\le L_{ll}   \]
hold. Moreover, the Cauchy-Schwarz inequality for the Hilbert-Schmidt inner product justifies the following chain of inequalities
\begin{eqnarray*}
 D_{jl} &=& \langle\cn_j(\pi_{\gr}),\cn_l(\pi_{\gr})\rangle_{HS}\\
&\le & ||\cn_j(\pi_{\gr})||_2 ||\cn_l(\pi_{\gr})||_2\\
&\le & \max\{||\cn_j(\pi_{\gr})||_2^2, ||\cn_l(\pi_{\gr})||_2^2  \}\\
&=& \max\{D_{jj}, D_{ll}\} .
\end{eqnarray*}
 Therefore, an application of Lemma \ref{matrix-lemma} allows us to conclude from (\ref{eq:convex-klesse-14}) that
\begin{eqnarray*}
\mathbb{E} (F_{c,e}(U\pi_{\fr}U^{\ast},\cn))&\ge& \textrm{tr}(\cn (\pi_{\gr}))\nonumber\\
&& -2\sum_{j=1}^{N}\sqrt{kn_j}||\cn_j(\pi_{\gr})||_2,
\end{eqnarray*}
which is what we aimed to prove.
\begin{flushright}$\Box$\end{flushright}
\section{\label{sec:finite-direct-part}Direct Part of The Coding Theorem: Finite Case}
In this section we will prove the coding theorem for finite compound channels. This is done in subsection \ref{subsec:finite-direct-part} after recalling some well-known properties of frequency-typical subspaces and Kraus operators in subsections \ref{subsec:typical-projections} and \ref{subsec:typical-kraus-operators} we
\subsection{\label{subsec:typical-projections}Typical Projections}
In this subsection we will collect some well-known results on typical projections.\\
Let $\rho\in \cs (\hr)$ be a state and consider any diagonalization
\[\rho=\sum_{i=1}^{d}\lambda_i|e_{i}\rangle\langle e_i| , \]
where $d:=\dim \hr$. Using this representation the state $\rho^{\otimes l}$ can be written as
\[\rho^{\otimes l}=\sum_{x^l\in A^l}\lambda_{x^l}|e_{x^l}\rangle\langle e_{x^l}|,  \]
with $A=\{1,\ldots, d  \}$, $x^l:=(x_1,\ldots, x_l)\in A^l$, $\lambda_{x^l}:=\lambda_{x_1}\cdot\ldots\cdot\lambda_{x_l}$, and $e_{x^l}:=e_{x_1}\otimes \ldots \otimes e_{x_l}$.\\
The frequency-typical set of eigenvalues of $\rho$ is given by
\[T_{\delta,l}:=\{x^l\in A^l: ||p_{x^l}-\lambda||_1<\delta, \  p_{x^l}\ll\lambda \},  \]
where $p_{x^l}$ denotes the empirical probability distribution on $A$ generated by $x^l$, i.e.
\[ p_{x^l}(x):=\frac{|\{j\in \{1,\ldots,l\}: x_j=x  \}  |}{l}, \]
$\lambda$ is the probability distribution on $A$ defined by the eigenvalues of $\rho$, and $p_{x^l}\ll\lambda $ means that $p_{x^l}(x)=0$ whenever $\lambda_x=0$.\\
The frequency-typical projection $q_{\delta,l}$ of $\rho$ given by
\[q_{\delta,l}:=\sum_{x^l\in T_{\delta,l}}|e_{x^l}\rangle\langle e_{x^l}|\]
has the following well-known properties:
\begin{lemma}\label{lemma-typical-1}
There is a real number $c>0$ and a function $h:\mathbb N\rightarrow\mathbb R_+$ with $h>0$ and $\lim_{l\rightarrow\infty}h(l)=0$ such that for each $ \delta\in (0,1/2)$ there exists a number $\varphi(\delta)>0$, with $\lim_{\delta\to 0}\varphi(\delta)= 0$, and for any $\rho\in \cs (\hr)$ the frequency-typical projection $q_{\delta,l}\in \mathcal{B}(\hr)^{\otimes l}$ satisfies
\begin{enumerate}
\item $\textrm{tr}(\rho^{\otimes l}q_{\delta,l})\ge 1-2^{-l(c\delta^2-h(l))}$,
\item $2^{-l(S(\rho)+\varphi(\delta) )}q_{\delta,l}\le q_{\delta,l}\rho^{\otimes l}q_{\delta,l}\le 2^{-l(S(\rho)-\varphi(\delta) )}q_{\delta,l}  $, and
\item $\eta_l(\delta)2^{ l(S(\rho)-\varphi(\delta))}\le\textrm{tr}(q_{\delta,l})\le 2^{l(S(\rho)+\varphi(\delta))}$ where
\[ \eta_l(\delta):= 1-2^{-l(c\delta^2-h(l))}.\]
\end{enumerate}
The rightmost inequalities in 2. and 3. imply
\[||q_{\delta,l}\rho^{\otimes l}q_{\delta,l}||_2^2 \le 2^{-l(S(\rho)-3\varphi(\delta))}. \]
Moreover, $\vphi(\delta)$ and $h$ are given by
\[h(l)=\frac{d}{l}\log(l+1)\ \ \forall l\in\mathbb N\]
\[ \vphi(\delta)=-\delta\log \frac{\delta}{d}. \]
\end{lemma}
The proof of the lemma is fairly standard and rests on purely classical reasoning. It combines the Bernstein-Sanov trick (cf. \cite{shields}, sect. III.1) and the type counting methods from \cite{csiszar}.
\subsection{\label{subsec:typical-kraus-operators}Typical Kraus Operators}
According to Kraus' representation theorem we can find to any $\cn\in \mathcal{C}(\hr,\kr) $ a family of operators $a_1,\ldots,a_n\in \mathcal{B}(\hr, \kr)$ with $\sum_{i=1}^{n}a_{i}^{\ast}a_i=\idn_{\hr}$ and
\[\cn(\rho)=\sum_{i=1}^{n}a_i\rho a_{i}^{\ast}  \]
for all $\rho\in\cs (\hr)$.\\
We fix the maximally mixed state $\pi_{\gr}$ supported by the subspace $\gr$ of $\hr$. It is easily seen (cf. \cite{nielsen}) that the Kraus operators $a_1,\ldots, a_n$ of $\cn$ can always be chosen such that
\[\textrm{tr}(a_i\pi_{\gr} a_j^{\ast} )=\delta_{ij}\textrm{tr}(a_i\pi_{\gr} a_i^{\ast}),  \]
for all $i,j\in\{1,\ldots, n  \}$. With this choice of Kraus operators we can define a probability distribution $r$ on the set $B:=\{1,\ldots, n  \}$ by
\[r(i):=\textrm{tr}(a_i\pi_{\gr} a_i^{\ast}),\qquad (i\in B).  \]
It is shown in \cite{schumacher} that the Shannon entropy of $r$ is nothing else than the entropy exchange $S_{e}(\pi_{\gr},\cn)$, i.e.
\[H(r)=S_{e}(\pi_{\gr}, \cn).  \]
In a similar vein as in the previous subsection we introduce the notion of frequency-typical subset for $r$, i.e we set
\[K_{\delta,l}:=\{y^l\in B^l: ||p_{y^l}-r||_1<\delta, \ p_{y^l}\ll r \}     \]
with $\delta>0$. With this we can introduce the notion of the reduced operation by setting
\begin{equation}\label{reduced-operation}
\cn_{\delta,l}(\rho):=\sum_{y^l\in K_{\delta,l}}a_{y^l}\rho a_{y^l}^{\ast},
\end{equation}
where $a_{y^l}:=a_{y_1}\otimes \ldots \otimes a_{y_l}$ and $\rho\in \cs (\hr^{\otimes l})$. Moreover, we set
\[n_{\delta,l}:=|K_{\delta,l}|,  \]
which is the number of Kraus operators of the reduced operation $\cn_{\delta,l}$.
The properties of frequency-typical sets (cf. \cite{csiszar, shields})) lead immediately to
\begin{lemma}\label{lemma-typical-2}
Let $\delta\in (0,1/2)$, $l\in\nn$, and fix the maximally mixed state $\pi_{\gr}$ on the subspace $\gr$ of $\hr$. There is a real number $\gamma( \delta)>0$ with $\lim_{\delta\to 0}\gamma(\delta)=0$ such that for each $\cn\in \mathcal{C}(\hr,\kr)$ the reduced operation $\cn_{\delta,l}$ satisfies
\begin{enumerate}
\item $\textrm{tr}(\cn_{\delta,l}(\pi_{\gr}^{\otimes l}))=r^{\otimes l}(K_{\delta,l})\ge 1-2^{-l(c'\delta^2-h'(l))}$, with a universal positive constant $c'>0$,
\item $n_{\delta,l}\le 2^{l(S_{e}(\pi_{\gr},\cn)+\gamma(\delta))}$.
\end{enumerate}
The function $h':\mathbb N\rightarrow\mathbb R_+$ is given by $h'(l)=\frac{d^2}{l}\log(l+1)$.
\end{lemma}
\subsection{\label{subsec:finite-direct-part}The Direct Coding Theorem for Finitely Many Channels}
Let us consider a compound channel given by a finite set $\fri:=\{\cn_1,\ldots ,\cn_{N} \}\subset \mathcal{C}(\hr,\kr)$ and two subspaces $\mathcal{E}_l, \gr^{\otimes l}$ of $\hr^{\otimes l}$ with $\mathcal{E}_l\subset\gr^{\otimes l}\subset \hr^{\otimes l}$. Let $k_l:=\dim \mathcal{E}_l$ and consider the associated maximally mixed states $\pi_{\mathcal{E}_l}$ on $\mathcal{E}_l$ and  $\pi_{\gr}$ on $\gr$.\\
For $j\in\{1,\ldots,N  \}$ and states $\cn_j(\pi_{\gr})$ let $q_{j,\delta,l}\in \mathcal{B}(\kr)^{\otimes l}$ denote the frequency-typical projection of $\cn_j(\pi_\gr)$ defined in Sec. \ref{subsec:typical-projections} for $\delta>0$ and $l\in\nn$. Moreover, let $\cn_{j,\delta,l}$ be the reduced operation associated with $\cn_j(\pi_{\gr})$ for $j\in\{1,\ldots, N  \},\delta>0,$ and $l\in\nn$ as considered in Sec. \ref{subsec:typical-kraus-operators}.\\
To every projection $q$ on $\hr^{\otimes l}$, associate the operation $\mathcal{Q}\in\mathcal{C}^{\downarrow}(\hr^{\otimes l},\hr^{\otimes l})$ defined by $\mathcal{Q}(\ \cdot\ ):=q\ \cdot\ q$ and let $q^\bot$ denote the ortho-complement of $q$.\\
In the following, for the compound channel $\fri$ and every
$l\in\mathbb N$, we will consider the average code fidelity
\[
F_{c,e,d}(\pi_{\mathcal{E}_l},\fri):=\max_{\crr_{\cn_1},...,\crr_{\cn_N}}\frac{1}{N}\sum_{i=1}^NF_e(\pi_{\mathcal{E}_l},\crr_{\cn_i}\circ\cn_i^{\otimes
l})
\]
where it is understood that $\crr_{\cn_i}$ are recovery
operations.\\
Note that for every $\epsilon\in(0,1)$,
$F_{c,e,d}(\pi_{\mathcal{E}_l},\fri)\geq1-\epsilon$ implies the existence
of an $(l,k_l)$ code with informed decoder for $\fri$ that
satisfies $F_e(\pi_{\mathcal{E}_l},\crr_{\cn_i}\circ\cn_i)\geq1-N\cdot\epsilon$ for every
$i\in\{1,...,N\}$.\\
For given $\delta>0$, $j\in\{1,...,N\}$ and $l\in \nn$ we define
\begin{eqnarray*}
\hat\cn_j(\ \cdot\
):=\mathcal{Q}_{j,\delta,l}\circ\cn_{j,\delta,l}(\ \cdot \ )
\end{eqnarray*}
and, accordingly,
\[\hat\cn(\ \cdot\ ):=\frac{1}{N}\sum_{j=1}^{N}\hat\cn_j(\ \cdot \ ).  \]
We show that, for every $j\in\{1,...,N\}$,
\begin{eqnarray}\label{eq:direct-finite-1}
F_{c,e}(\pi_{\mathcal{E}_l},\cn_j^{\otimes l})\geq
F_{c,e}(\pi_{\mathcal{E}_l},\hat\cn_j).
\end{eqnarray}
Let $\crr \in\mathcal{C}(\kr^{\otimes l},\hr^{\otimes l})$ be any
recovery operation. Then
\begin{eqnarray}
F_{c,e}(\pi_{\mathcal{E}_l},\cn_j^{\otimes l})
&\geq&F_e(\pi_{\mathcal{E}_l},\crr\circ(\mathcal{Q}_{j,\delta,l}+\mathcal{Q}^\bot_{j,\delta,l})\circ\cn_j^{\otimes l})\nonumber\\
&=&F_e(\pi_{\mathcal{E}_l},\crr\circ\mathcal{Q}_{j,\delta,l}\circ\cn_j^{\otimes l})\nonumber\\
&&+F_e(\pi_{\mathcal{E}_l},\crr\circ\mathcal{Q}^\bot_{j,\delta,l}\circ\cn_j^{\otimes l})\nonumber\\
&\geq&F_e(\pi_{\mathcal{E}_l},\crr\circ\mathcal{Q}_{j,\delta,l}\circ\cn_j^{\otimes l})\nonumber\\
&\geq&F_e(\pi_{\mathcal{E}_l},\crr\circ\hat\cn_j)\nonumber.
\end{eqnarray}
The last inequality follows from the fact that $\hat\cn_j$ is a reduction of $\mathcal{Q}_{j,\delta,l}\circ\cn_j^{\otimes l}$. Taking the maximum over all recovery operations $\crr$ proves the claim.\\
Using (\ref{eq:direct-finite-1}), we get the lower bound
\begin{eqnarray}\label{eq:direct-finite-2}
F_{c,e,d}(\pi_{\mathcal{E}_l},\fri)&=&\frac{1}{N}\sum_{i=1}^NF_{c,e}(\pi_{\mathcal{E}_l},\cn_i^{\otimes l})\nonumber\\
&\geq&\frac{1}{N}\sum_{i=1}^NF_{c,e}(\pi_{\mathcal{E}_l},\hat\cn_i)\nonumber\\
&=&F_{c,e,d}(\pi_{\mathcal{E}_l},\hat\cn)\nonumber\\
&\geq&F_{c,e}(\pi_{\mathcal{E}_l},\hat\cn)
\end{eqnarray}
Note that the inequality (\ref{eq:direct-finite-2}) is still valid
if we replace $\pi_{\mathcal{E}_l}$ by $u\pi_{\mathcal{E}_l}u^{\ast}$ for any
$u\in \mathfrak{U}(\gr^{\otimes l})$. Consider
\[\bar{F}_{e}(\gr,\fri):=\mathbb{E}(F_{c,e,d}(U\pi_{\mathcal{E}_l}U^{\ast},\fri  ) )  \] with a random variable $U$ taking values in
$\mathfrak{U}(\gr^{\otimes l})$ and which is distributed according
to the Haar measure. From (\ref{eq:direct-finite-2}) it follows
that
\begin{equation}\label{eq:direct-finite-3}
  \bar{F}_{e}(\gr,\fri)\ge \mathbb{E}(F_{c,e}(U\pi_{\mathcal{E}_l}U^{\ast},\hat\cn  ) ). \end{equation}
Applying Theorem \ref{convex-klesse} to the right hand side of
(\ref{eq:direct-finite-3}) we arrive at
\begin{eqnarray}\label{eq:direct-finite-4}
\bar{F}_{e}(\gr,\fri)  &\ge&   \mathbb{E}(F_{c,e}(U\pi_{\mathcal{E}_l}U^{\ast},\hat\cn)  )\nonumber\\
&\ge& \textrm{tr}(\hat\cn(\pi_{\gr}^{\otimes l}) )\nonumber\\
&& -2\sum_{j=1}^{N}\sqrt{k_ln_{j,\delta,l}}||\hat\cn_{j}(\pi_{\gr}^{\otimes l})||_2,
\end{eqnarray}
where $n_{j,\delta,l}$ stands for the number of Kraus operators of the reduced operation $\hat\cn_j$.
\begin{theorem}[Direct Part: Informed Decoder]\label{lemma-direct-finite-1}\
Let $\fri=\{\cn_1,...,\cn_N\}\subset \mathcal{C}(\hr,\kr)$ be a compound channel and $\pi_\gr$ the maximally mixed state associated to a subspace $\gr\subset\hr$. Then
$$Q_{ID}(\fri)\geq\min_{\cn_i\in\fri}I_c(\pi_\gr,\cn_i).$$
\end{theorem}
\emph{Remark}. Arguing as Klesse \cite{klesse}, an immediate consequence of this theorem is the inequality
$$Q_{ID}(\fri)\geq\lim_{l\rightarrow\infty}\frac{1}{l}\max_{\rho\in\mathcal{S}(\hr^{\otimes l})}\min_{\cn_i\in\fri}I_c(\rho,\cn_i^{\otimes l}).$$
See also the argument in section \ref{sec:direct-part}.
Using the quantum Fano inequality and following the lines of \cite{bkn}, it is easy to establish the converse
\begin{equation}\label{eq:converse-finite}
Q_{ID}(\fri)\leq\lim_{l\rightarrow\infty}\frac{1}{l}\max_{\rho\in\mathcal{S}(\hr^{\otimes l})}\min_{\cn_i\in\fri}I_c(\rho,\cn_i^{\otimes l}).
\end{equation}
\emph{Proof}. We show that for every $\epsilon>0$ the number $\min_{\cn_i\in\fri}I_c(\pi_\gr,\cn)-\epsilon$ is an achievable rate for $\fri$.\\
1) If $\min_{\cn_i\in\fri}I_c(\pi_\gr,\cn_i)-\epsilon\leq0$, there is nothing to prove.\\
2) Let $\min_{\cn_i\in\fri}I_c(\pi_\gr,\cn_i)-\epsilon>0$.\\
Choose $\delta\in(0,1/2)$ satisfying $\gamma(\delta)+3\vphi(\delta)<\epsilon$ with functions $\psi,\vphi$ from Lemma \ref{lemma-typical-1}, \ref{lemma-typical-2}.\\
Pick a sequence of subspaces $(\pi_{\mathcal{E}_l})_{l\in\mathbb N}$ with $\pi_{\mathcal{E}_l}\leq\pi_\gr^{\otimes l}$ for all $l\in\mathbb N$ that are of dimension
$k_l=\lfloor2^{l(\min_{\cn_i\in\fri}I_c(\pi_\gr,\cn_i)-\epsilon)}\rfloor$. By $S(\pi_\gr)\geq I_c(\pi_\gr,\cn_j)$ (see \cite{bkn}), this is always possible.\\
Obviously, $\frac{1}{l}\log k_l\leq\min_{\cn_i\in\fri}I_c(\pi_\gr,\cn_i)-\epsilon$.\\
We consider the terms in (\ref{eq:direct-finite-3}) separately.
\begin{eqnarray}\label{eq:direct-finite-4}
\textrm{tr}(\hat\cn(\pi_{\gr}^{\otimes l}) )&=&\frac{1}{N}\sum_{j=1}^N[\mathrm{tr}(\mathcal{Q}_{j,\delta,l}\circ\cn_j^{\otimes l}(\pi_{\gr}^{\otimes l}))\nonumber\\
&&+\mathrm{tr}(\mathcal{Q}_{j,\delta,l}\circ(\cn_{j,\delta,l}-\cn_j^{\otimes l})(\pi_{\gr}^{\otimes l})]\nonumber\\
&\geq&\frac{1}{N}\sum_{j=1}^N[\mathrm{tr}(\mathcal{Q}_{j,\delta,l}\circ\cn_j^{\otimes l}(\pi_{\gr}^{\otimes l}))\nonumber\\
&&+\mathrm{tr}((\cn_{j,\delta,l}-\cn_j^{\otimes l})(\pi_{\gr}^{\otimes l})]\nonumber\\
&=&\frac{1}{N}\sum_{j=1}^N[\mathrm{tr}(q_{j,\delta,l}\cn_j^{\otimes l}(\pi_{\gr}^{\otimes l}))\nonumber\\
&&-\mathrm{tr}(\sum_{y^l\notin K_{j,\delta,l}}a_{j,y^l}\pi_{\gr}^{\otimes l}a_{j,y^l}^*]\nonumber\\
&\geq&1-2^{-l(c\delta^2-h(l))}-2^{-l(c'\delta^2-h'(l))}.
\end{eqnarray}
We used the fact that $\cn_{j,\delta,l}$ is a reduction of $\cn_j^{\otimes l}$ and Lemmas \ref{lemma-typical-1},\ref{lemma-typical-2}.\\
Further, using the inequality $||A+B||_2^2\geq||A||_2^2+||B||_2^2$ valid for nonnegative operators $A,B\in\mathcal{B}(\kr^{\otimes l})$ (see \cite{klesse}), we get the inequality
\begin{eqnarray}\label{eq:direct-finite-5}
||\hat\cn_j(\pi_{\gr}^{\otimes l})||_2^2&\leq&||q_{j,\delta,l}\cn_{j,\delta,l}(\pi_{\gr}^{\otimes l})||_2^2\nonumber\\
&&+||q_{j,\delta,l}(\cn_j^{\otimes l}-\cn_{j,\delta,l})(\pi_{\gr}^{\otimes l})||_2^2\nonumber\\
&\leq&||q_{j,\delta,l}(\cn_{j,\delta,l}+[\cn_j^{\otimes l}-\cn_{j,\delta,l}])(\pi_{\gr}^{\otimes l})||_2^2\nonumber\\
&=&||q_{j,\delta,l}\cn_j^{\otimes l}(\pi_{\gr}^{\otimes l})||_2^2\nonumber\\
&=&\mathrm{tr}(q_{j,\delta,l}\cn_j^{\otimes l}(\pi_{\gr}^{\otimes l})q_{j,\delta,l}^2\cn_j^{\otimes l}(\pi_{\gr}^{\otimes l})q_{j,\delta,l})\nonumber\\
&\leq&\mathrm{tr}(q_{j,\delta,l})2^{-2l(S(\cn_j^{\otimes l}(\pi_{\gr}^{\otimes l}))-\vphi(\delta))}\nonumber\\
&\leq&2^{-l(S(\cn_j^{\otimes l}(\pi_{\gr}^{\otimes l}))-3\vphi(\delta))}.
\end{eqnarray}
The last two inequalities follow from Lemma \ref{lemma-typical-1}. Combining the inequalities (\ref{eq:direct-finite-3}), (\ref{eq:direct-finite-4}) and (\ref{eq:direct-finite-5}) with Lemma \ref{lemma-typical-2}, we finally see that
\begin{eqnarray}
\bar{F}_{e}(\gr,\fri)&\ge&1-2^{-l(c\delta^2-h(l))}-2^{-l(c'\delta^2-h'(l))}\nonumber\\
&&-2\sum_{j=1}^N\sqrt{2^{l(\frac{1}{l}\log k_l+\gamma(\delta)+3\vphi(\delta)-I_c(\pi_{\gr},\cn_j)}}\nonumber\\
&\geq&1-2^{-l(c\delta^2-h(l))}-2^{-l(c'\delta^2-h'(l))}\nonumber\\
&&-2N\sqrt{2^{-l(\epsilon-\gamma(\delta)-3\vphi(\delta))}}\nonumber.
\end{eqnarray}
This shows the existence of at least one sequence of $(l,k_l)$ codes for $\fri$ with informed decoder and
\[ \lim_{l\rightarrow\infty}\frac{1}{l}\log k_l=\min_{\cn_i\in\fri}I_c(\pi_{\gr},\cn_i)-\epsilon \]
as well as, for every $l\in\mathbb N$,
\[ \min_{\cn\in\fri}F_e(\pi_{\fr_l},\crr_\cn\circ\cn^{\otimes l})\geq1-N\cdot\epsilon_l \]
with
\[\epsilon_l=2^{-l(c\delta^2-h(l))}+2^{-l(c'\delta^2-h'(l))}+2N\sqrt{2^{-l(\epsilon-\gamma(\delta)-3\vphi(\delta))}}.\]
\begin{flushright}$\Box$\end{flushright}
Our next goal is to convert the codes for the given, finite set $\fri\subset\mathcal{C}(\hr,\kr)$ with the informed decoder into truly compound codes. To this end, we need the channel estimation technique of Datta and Dorlas \cite{datta-dorlas} which we recall now:
\begin{theorem}[Datta \& Dorlas \cite{datta-dorlas}]\label{datta-dorlas-theorem}\ \\
Let $\fri=\{\cn_1,\ldots \cn_N  \}\subset\mathcal{C}(\hr,\kr)$ with $N\in\mathbb N$. Set $L={N \choose 2}$. There is $f=f(\fri)\in (0,1)$ such that for each $m\in\nn$ we can find mutually orthogonal projections $p_{1,mL},\ldots p_{N,mL}\in\mathcal{B}(\kr^{\otimes mL})$ with $\sum_{i=1}^{N}p_{i,mL}=\idn_{\hr}^{\otimes mL}$ and a pure state $\omega^{(mL)}\in\mathcal{S}(\hr^{\otimes mL})$ such that
\[ \textup{tr}(p_{i,mL} \cn_i^{\otimes mL} (\omega^{(mL)}))\ge (1-N f^m)^{N-1} \]
holds for all $i\in\{1,\ldots, N  \} $.
\end{theorem}
With the help of theorem \ref{datta-dorlas-theorem} we will show that knowledge of the channel in the finite compound scenario does not increase the capacity.
\begin{lemma}\label{conversion-of-codes}
For any finite set $\fri=\{\cn_1,\ldots \cn_N  \}\subset\mathcal{C}(\hr,\kr) $ we have
\[ Q_{ID}(\fri)=Q(\fri). \]
\end{lemma}
\emph{Proof.} The inequality $Q_{ID}(\fri)\ge Q(\fri)$ is obvious from the definitions of these capacities.\\
The proof will be complete if we show that each rate $R$ achievable within the scenario of an informed decoder is also achievable without knowledge of the channel. Let $(\fr_t, \{ \crr_i^t \}_{i=1}^N)$ be a $(t,k_t)$-code for $\fri=\{ \cn_1,\ldots,\cn_N \}$ with informed decoder and
\begin{equation}\label{eq:conversion-of-codes-1}
 F_e(\pi_{\fr_t}, \crr_i^t\circ \cn_i^{\otimes t} )\ge 1-\eps
\end{equation}
for some $\eps\in (0,1)$  and all $i\in\{1,\ldots, N  \}$.\\
Let the pure state $\omega^{(mL)}$ and projections $p_{i,mL},\ldots,p_{N,mL}$ be as described in Theorem \ref{datta-dorlas-theorem}.
Take $x_{mL}\in\hr^{\otimes mL}$ such that $\omega^{(mL)}=|x_{mL}\rangle\langle x_{mL}|$ and define a set of measurement operations $\hat{R}_i^{mL}\in \mathcal{C}^{\downarrow}(\kr^{\otimes mL},\hr^{\otimes mL})$ by $\hat{R}_i^{mL}(\ \cdot\ ):=\omega^{(mL)}\textup{tr}(p_{i,mL}\ \cdot\ p_{i,mL})$.\\
Set
\[\crr^{mL+t}:=\sum_{i=1}^{N}\hat{R}_i^{mL}\otimes \crr_i^t,  \]
\[\fr'_{mL+t}:=\{\cc\cdot x_{mL}  \}\otimes \fr_{t}  \]
and consider the averaged channel
\[ \cn^{mL+t}=\frac{1}{N}\sum_{i=1}^{N}\cn_i^{\otimes (mL+t)}.\]
With the abbreviation
\[F_e:= F_{e}(\pi_{\fr'_{mL+t}}, \crr^{mL+t}\circ \cn^{mL+t}) \]
we arrive at
\begin{eqnarray}\label{eq:conversion-of-codes-2}
  F_{e}&=& \frac{1}{N}\sum_{i=1}^{N}F_e(\pi_{\fr'_{mL+t}}, \crr^{mL+t}\circ \cn_i^{\otimes (mL+t)} )\nonumber\\
&=&\frac{1}{N}\sum_{i,j=1}^{N}F_e(\omega^{(mL)}\otimes \pi_{\fr_t},(\hat{R}_i^{mL}\otimes \crr_j^t)\circ \cn_i^{\otimes (mL+t)} )\nonumber\\
&\ge& \frac{1}{N}\sum_{i=1}^{N}F_e(\omega^{(mL)},\hat{R}_i^{mL}\circ\cn_i^{\otimes mL}  )F_e(\pi_{\fr_t}, \crr_i^t\circ \cn_i^{\otimes t} )\nonumber\\
&\ge& \frac{1}{N}\sum_{i=1}^{N}\langle x_{mL},\omega^{(mL)}\textup{tr}(p_{i,mL}\cn_i(\omega^{(mL)})) x_{mL}\rangle(1-\eps)\nonumber\\
&\ge& (1-N f^m)^{N-1}(1-\eps),
\end{eqnarray}
by theorem \ref{datta-dorlas-theorem} and (\ref{eq:conversion-of-codes-1}). We see from (\ref{eq:conversion-of-codes-2}) that for each $i\in\{1,\ldots, N  \}$
\begin{equation}\label{eq:conversion-of-codes-3}
 F_{e}(\pi_{\fr'_{mL+t}}, \crr^{mL+t}\circ \cn_i^{\otimes (mL+t)})\geq 1-N(1-\sigma(N,m,\eps,f) )
\end{equation}
with $\sigma(N,m,\eps,f)=(1-N f^m)^{N-1}(1-\eps) $.
Replacing, for sufficiently large $l\in\nn$, $m$ by $\lfloor \sqrt{l}\rfloor$, $t$ by $l-L\lfloor \sqrt{l}\rfloor$, and $\eps$ by a sequence $(\eps_t)_{t\in\nn}$ with $\lim_{t\to\infty}\eps_t=0$ we obtain a sequence of $(l,k_l)$-codes $(\fr'_l,\crr^l)$ for $\fri$ with
\[\lim_{l\to\infty}\min_{i\in\{1,\ldots, N  \}}F_e(\pi_{\fr'_l}, \crr^l\circ \cn_i^{\otimes l})=1,  \]
and
\[\liminf_{l\to\infty}\frac{1}{l}\log \dim \fr'_l =\liminf_{t\to\infty}\frac{1}{t}\log k_t, \]
which concludes the proof.
\begin{flushright}$\Box$\end{flushright}
Lemma \ref{conversion-of-codes}, Theorem \ref{lemma-direct-finite-1}, and (\ref{eq:converse-finite}) immediately imply the following coding theorem.
\begin{theorem}[Coding Theorem: $|\fri|<\infty$]
For any finite set $\fri=\{\cn_1,\ldots \cn_N  \}\subset\mathcal{C}(\hr,\kr) $ we have
\[Q(\fri)=\lim_{l\rightarrow\infty}\frac{1}{l}\max_{\rho\in\mathcal{S}(\hr^{\otimes l})}\min_{i\in\{1,\ldots, N  \}}I_c(\rho,\cn_i^{\otimes l}).\]
\end{theorem}
Utilizing lemma \ref{finite-averaged-vs-compound} we obtain immediately the following capacity result for finite  averaged quantum channels.
\begin{corollary}[Capacity of Averaged Channels]
For any finite set $\fri=\{\cn_1,\ldots \cn_N  \}\subset\mathcal{C}(\hr,\kr) $ and any $\lambda=(\lambda_1,\ldots \lambda_N)\in \rr^{N}$ with $\lambda_1,\ldots,\lambda_N>0$, $\sum_{i=1}^{N}\lambda_i=1$ we have
\[Q(\mathfrak{A}_{\fri,\lambda} )=\lim_{l\rightarrow\infty}\frac{1}{l}\max_{\rho\in\mathcal{S}(\hr^{\otimes l})}\min_{i\in\{1,\ldots, N  \}}I_c(\rho,\cn_i^{\otimes l}),  \]
where $\mathfrak{A}_{\fri,\lambda} $ denotes the averaged channel defined by $\fri$ and $\lambda$.
\end{corollary}
We shall now briefly discuss the case of an informed encoder for finite compound channels.
\begin{lemma}\label{existence-of-subcodes}
Let $\pi_C$ be the maximally mixed state on a subspace $C\subset\hr^{\otimes l}$ of dimension $D$ and $\mathcal E:\mathcal{B}(\hr^{\otimes l})\rightarrow\mathcal{B}(\hr^{\otimes l})$ such that
$$F_e(\pi_C,\mathcal E)\geq1-\epsilon.$$
To every $K\leq D$ there exists a subcode $C'\subset C$ of dimension $\dim C'=K$ such that
$$F_e(\pi_{C'},\mathcal E)\geq1-\frac{D}{\lfloor\frac{D}{K}\rfloor\cdot K}\epsilon.$$
\end{lemma}
\emph{Proof.} Let $P$ be the orthogonal projection onto $C$, that is $\pi_C=\frac{1}{\textup{tr}(P)}P$. Set $L:=\lfloor\frac{D}{K}\rfloor$. Take a decomposition
$$C=\bigoplus_{i=1}^{L}C_i\bigoplus C_{L+1}$$
of $C$ into pairwise orthogonal subspaces $C_i\subset C$ that satisfies $\dim C_i=K,\ i\in\{1,...,L\}$ and $\dim C_{L+1}=D-K\cdot L$.\\
To every $C_i$, let the maximally mixed state on $C_i$ be denoted by $\pi_{C_i}$. Clearly,
$$\pi_C=\sum_{i=1}^{L}\frac{K}{D}\pi_{C_i}+\frac{D-K\cdot L}{D}\pi_{C_{L+1}}.$$
By assumption and use of convexity of entanglement fidelity in the input distribution, this implies
\begin{eqnarray}
1-\epsilon&\leq&F_e(\pi_C,\mathcal E)\nonumber\\
&\leq&\sum_{i=1}^{L}\frac{K}{D}F_e(\pi_{C_i},\mathcal E)+\frac{D-K\cdot L}{D}F_e(\pi_{C_{L+1}},\mathcal E)\nonumber\\
&\leq&L\frac{K}{D}\max_{1\leq i\leq L\cdot K}F_e(\pi_{C_i},\mathcal E))+1-L\frac{K}{D}.\label{existence of good subcode}
\end{eqnarray}
Clearly, (\ref{existence of good subcode}) implies that there exists $j\in\{1,...,L\frac{K}{D}\}$ such that
$$1-\frac{D}{L\cdot K}\epsilon\leq F_e(\pi_{C_j},\mathcal E).$$
Set $C'=C_j$.\begin{flushright}$\Box$\end{flushright}
\begin{lemma}\label{scaling-of-exponentially-large-subspaces}
Let $n\in\mathbb N$, $A,B\in\mathbb R$ and $A>B>0$. Then
$$\frac{\lfloor2^{nA}\rfloor}{\lfloor2^{nB}\rfloor\cdot\lfloor\frac{\lfloor2^{nA}\rfloor}{\lfloor2^{nB}\rfloor}\rfloor}\leq1-3\cdot2^{-nB}.$$
\end{lemma}
\emph{Proof.} Let, for the moment, $n$ be fixed. There are $\delta_A,\delta_B,\delta_{AB}\in\mathbb(0,1)$ such that $\lfloor2^{nA}\rfloor=2^{nA}+\delta_A$, $\lfloor2^{nB}\rfloor=2^{nB}+\delta_B$ and $\lfloor\frac{\lfloor2^{nA}\rfloor}{\lfloor2^{nB}\rfloor}\rfloor=\frac{2^{nA}+\delta_A}{2^{nB}+\delta_B}+\delta_{AB}$.
We will now derive a lower bound on the denominator.
\begin{eqnarray}
(2^{nB}+\delta_B)(\frac{2^{nA}+\delta_A}{2^{nB}+\delta_{B}}+\delta_{AB})&\geq&2^{nB}\frac{2^{nA}}{2^{nB}+\delta_B}\nonumber\\
&=&2^{n(A+B)}\frac{1}{2^{nB}+\delta_B}\nonumber\\
&\geq&2^{n(A+B)}\frac{1}{2^{nB}+1}\label{erster bound}
\end{eqnarray}
With the help of (\ref{erster bound}) we get
\begin{eqnarray}
\frac{\lfloor2^{nA}\rfloor}{\lfloor2^{nB}\rfloor\cdot\lfloor\frac{\lfloor2^{nA}\rfloor}{\lfloor2^{nB}\rfloor}\rfloor}&=&\frac{2^{nA}+\delta_A}{(2^{nB}+\delta_B)(\frac{2^{nA}+\delta_A}{2^{nB}+\delta_{B}}+\delta_{AB})}\nonumber\\
&\leq&(2^{nA}+\delta_A)2^{-n(A+B)}(2^{nB}+1)\nonumber\\
&\leq&(2^{nA}+1)2^{-n(A+B)}(2^{nB}+1)\nonumber\\
&=&(2^{-nB}+2^{-n(A+B)})(2^{nB}+1)\nonumber\\
&=&1+2^{-nB}+2^{-nA}+2^{-n(A+B)}\nonumber\\
&\leq&1+3\cdot2^{-nB}\nonumber.
\end{eqnarray}
\begin{flushright}$\Box$\end{flushright}
\begin{theorem}[Converse: Informed Encoder]
Let $\fri=\{\cn_1,\ldots,\cn_N\}\subset\mathcal{C}(\hr,\kr)$ be a finite compound channel. The quantum capacity with informed encoder of $\fri$ is bounded from above by
$$Q_{IE}(\fri)\leq\min_{\cn_j\in\fri}Q(\cn_j).$$
\end{theorem}
This is easily seen using standard techniques from \cite{bkn}.
\\\\
We will now show that our upper bound is indeed achievable. To this end, we will now convert codes for single channels into codes for $\fri$ with informed encoder. Again, we need the channel estimation technique of Datta and Dorlas \cite{datta-dorlas} which has already been stated in Theorem \ref{datta-dorlas-theorem}.
\begin{theorem}[Direct Part: Informed Encoder]\label{conversion-of-codes}
For any finite set $\fri=\{\cn_1,\ldots \cn_N  \}\subset\mathcal{C}(\hr,\kr) $ we have
\[ Q_{IE}(\fri)\geq\min_{\cn_i\in\fri}Q(\cn_i). \]
\end{theorem}
\emph{Proof.} Throughout the proof, we will write 'for all $i$' meaning 'for all $i\in\{1,...,N\}$'.\\
Let $\epsilon>0$ be arbitrary and $\{\pi_{\gr_1},\ldots\pi_{\gr_N}\}$ be maximally mixed states on corresponding subspaces $\gr_i\subset\hr$. By the noisy channel coding theorem \cite{klesse}, for every $i$ there are sequences $(\fr_l^i,\crr_i^l)_{l\in\mathbb N}$, $(\epsilon_l^{(i)})_{l\in\mathbb N}$ with $\epsilon_l^{(i)}\searrow0$ such that
\[F_e(\pi_{\fr_l^i},\crr_i^l\circ\cn_i^{\otimes l})\geq1-\epsilon_l\]
$$\dim\fr_l^i=\lfloor2^{l(I_c(\pi_{\gr_i},\cn_i)-\epsilon)}\rfloor,$$
where $\epsilon_l:=\max_{1\leq i\leq N}\epsilon_l^{(i)}$ and, clearly, $\epsilon_l\searrow0$.\\
We set $E:=\{i|I_c(\pi_{\gr_i},\cn_i)=\min_{1\leq j\leq N}I_c(\pi_{\gr_j},\cn_j)\}$ and denote the complement of $E$ within $\{1,...,N\}$ by $E^\complement$.
We take $k:=\min\{i|i\in E\}$.\\
By definition, for every $j\in E^\complement$ we have $I_c(\pi_{\gr_k},\cn_k)-\epsilon<I_c(\pi_{\gr_j},\cn_j)-\epsilon$.
By lemmas \ref{existence-of-subcodes} and \ref{scaling-of-exponentially-large-subspaces}, for $j\in E^\complement$ there exist subspaces $\hat\fr_l^j$ with $\dim\hat\fr_l^j=\lfloor2^{(I_c(\pi_{\gr_k},\cn_k)-\epsilon}\rfloor$ such that
$$F_e(\pi_{\hat\fr_l^j},\crr_l^j\circ\cn_j^{\otimes l})\geq1-\epsilon_l(1+3\cdot2^{-l(I_c(\pi_{\gr_k},\cn_k)-\epsilon)}).$$
Setting, for every $i\in E$, $\hat\fr_l^i:=\fr_l^i$ and using the abbreviation $\delta_l:=\epsilon_l(1+3\cdot2^{-l(I_c(\pi_\gr,\cn_k)-\epsilon)})$ -observe that $\delta_l\searrow0$- we get
\begin{eqnarray}F_e(\pi_{\fr_l^i},\crr_i^l\circ\cn_i^{\otimes l})\geq1-\delta_l,\label{uniform-convergence-of-F_e}\end{eqnarray}
$$\lim_{l\rightarrow\infty}\frac{1}{l}\log\dim\fr_l^i=I_c(\pi_{\gr_k},\cn_k).$$
Let $m,t\in\mathbb N$ be arbitrary and let the pure state $\omega^{(mL)}$ and projections $p_{i,mL},\ldots,p_{N,mL}$ be as described in theorem \ref{datta-dorlas-theorem}.
Take $x_{mL}\in\hr^{\otimes mL}$ such that $\omega^{(mL)}=|x_{mL}\rangle\langle x_{mL}|$ and define a set of measurement operations $\hat{R}_i^{mL}\in \mathcal{C}^{\downarrow}(\kr^{\otimes mL},\hr^{\otimes mL})$ by $\hat{R}_i^{mL}(\ \cdot\ ):=\omega^{(mL)}\textup{tr}(p_{i,mL}\ \cdot\ p_{i,mL})$.\\
Set
\[\crr^{mL+t}:=\sum_{i=1}^{N}\hat{R}_i^{mL}\otimes \crr_i^t,  \]
\[\fr^{'i}_{mL+t}:=\{\cc\cdot x_{mL}  \}\otimes \hat\fr_{t}^i.  \]
With the abbreviation
\[F_e^i:= F_{e}(\pi_{\fr^{'i}_{mL+t}}, \crr^{mL+t}\circ \cn^{\otimes(mL+t)}) \]
we arrive at
\begin{eqnarray}\label{eq:conversion-of-codes-2}
F_{e}^i&=&\sum_{j=1}^{N}F_e(\omega^{(mL)},\hat{R}_j^{mL}\circ\cn_i^{\otimes mL}  )F_e(\pi_{\hat\fr_t^i}, \crr_j^t\circ \cn_i^{\otimes t} )\nonumber\\
&\ge&F_e(\omega^{(mL)},\hat{R}_i^{mL}\circ\cn_i^{\otimes mL}  )F_e(\pi_{\hat\fr_t^i}, \crr_i^t\circ \cn_i^{\otimes t} )\nonumber\\
&\ge&\langle x_{mL},\omega^{(mL)}\textup{tr}(p_{i,mL}\cn_i(\omega^{(mL)})) x_{mL}\rangle(1-\delta_t)\nonumber\\
&\ge&(1-N f^m)^{N-1}(1-\delta_t),
\end{eqnarray}
Replacing, for sufficiently large $l\in\nn$, $m$ by $\lfloor \sqrt{l}\rfloor$, $t$ by $l-L\lfloor \sqrt{l}\rfloor$, we obtain a sequence of $(l,k_l)$-codes for $\fri$ with
informed encoder that achieves
$$\min_{1\leq j\leq N}I_c(\pi_{\gr_j},\cn_j)-\epsilon.$$
Since this holds true for every $\epsilon>0$ we know that for every set $\{\pi_{\gr_1},\ldots,\pi_{\gr_N}\}$ the number $\min_{1\leq j\leq N}I_c(\pi_{\gr_j},\cn_j)$ is an achievable rate.\\
Let $\{\sigma^{(m)}_i\}_{1\leq i\leq N}$ be such that $I_c(\sigma_i^{(m)},\cn_i^{\otimes m})=\max_{\rho^{(m)}}I_c(\rho^{(m)},\cn_i^{\otimes m})$ for all $i$ and let $(\pi_{n,\epsilon,i}^{(m)})_{n\in\mathbb N}$ be the sequence of frequency typical states associated to $(\sigma_i^{(m)\otimes n})_{n\in\mathbb N}$. We get the following chain of inequalities.
\begin{eqnarray*}
Q_{IE}(\fri)&\overset{(1)}{=}&\frac{1}{m}\lim_{n\rightarrow\infty}\frac{1}{n}Q_{IE}(\fri^{\otimes nm})\\
&\overset{(2)}{\geq}&\frac{1}{m}\lim_{\epsilon\rightarrow0}\lim_{n\rightarrow\infty}\frac{1}{n}\min_{1\leq j\leq N}I_c(\pi^{(m)}_{n,\epsilon,j},(\cn_j^{\otimes m})^{\otimes n})\\
&\overset{(3)}{=}&\min_{1\leq j\leq N}\frac{1}{m}\lim_{\epsilon\rightarrow0}\lim_{n\rightarrow\infty}\frac{1}{n}I_c(\pi^{(m)}_{n,\epsilon,j},(\cn_j^{\otimes m})^{\otimes n})\\
&\overset{(4)}{=}&\min_{1\leq j\leq N}\frac{1}{m}I_c(\sigma^{(m)}_j,\cn_j^{\otimes m})\\
&\overset{(5)}{=}&\min_{\cn_j\in\fri}\frac{1}{m}\max_{\rho^{(m)}}I_c(\rho^{(m)},\cn_j^{\otimes m}).
\end{eqnarray*}
Here, $(1)$ is easily seen using using the fact that pure states can be transmitted with entanglement fidelity one over any channel by using a constant recovery operation. $(2)$ has just been proven, while $(3)$ is just continuity of the function $\min$. In $(4)$, we use the variant of BSST lemma \ref{compound-bsst-lemma} for single channel. The last line $(5)$ follows by choice of $\sigma_i^{(m)}$.
Taking the limit $m\rightarrow\infty$ proves the claim.
\begin{flushright}$\Box$\end{flushright}
\section{\label{sec:nets}Nets and Discrete Approximation in the Set of Quantum Channels}
The purpose of this section is to collect and provide necessary building blocks for discretization arguments to follow in section \ref{sec:direct-part}. We start with the existence of good discrete approximation in the set $\mathcal{C}(\hr,\kr)$ with respect to the diamond norm $||\cdot||_{\lozenge}$ in subsection \ref{subsec:existence-nets}. The reasons for using this norm are that a) it is multiplicative with respect to tensor products of channels, and b) that it is easy to relate to the entanglement fidelity and coherent information. These relations are made precise in subsection \ref{subsec:compound-approx}.

\subsection{\label{subsec:existence-nets}Existence of Good Nets}
Let $\mathcal{C}(\hr,\kr)$ denote the set of CPTP maps with domain $\mathcal{B}(\hr)$ and range $\mathcal{B}(\kr)$. Without loss of generality we assume that $\hr=\cc^d$ and $\kr=\cc^{d'}$.
The constructions to follow will be based on the diamond norm on the set of channels. It is defined by the prescription
\begin{equation}\label{diamond-def}
||\mathcal{N}||_{\lozenge}:=\sup_{n\in \nn}\max_{a\in \mathcal{B}(\cc^n\otimes\hr),||a||_1=1}||(id_{n}\otimes \mathcal{N})(a)||_1 ,
\end{equation}
where $||\cdot ||_1$ stands for the trace norm, $id_n:\mathcal{B}(\cc^n)\to \mathcal{B}(\cc^n)$ is the identity channel, and $\mathcal{N}:\mathcal{B}(\hr)\to \mathcal{B}(\kr)$ is any linear map, not necessarily completely positive. $||\cdot||_{\lozenge}$ is the dual norm to $||\cdot||_{cb}$, the norm of complete boundedness. We refer to \cite{paulsen} and \cite{kitaev} for additional information on these norms. It is an important fact that we need not maximize over all non-negative integers in (\ref{diamond-def}), the norm stabilizes at the value $n=d=\dim \hr$, i.e.
\begin{equation}\label{diamond-stabil}
 ||\mathcal{N}||_{\lozenge}= \max_{a\in \mathcal{B}(\cc^d\otimes\hr),||a||_1=1}||(id_{d}\otimes \mathcal{N})(a)||_1.
\end{equation}
Two different proofs of this fact can be found in \cite{paulsen} and \cite{kitaev}.\\
It is well known, and not difficult to show, that for $\mathcal{N}\in \mathcal{C}(\hr,\kr)$ the relation
\[||\cn||_{\lozenge}=1  \]
holds. Consequently, $\mathcal{C}(\hr,\kr)\subset \{\Phi:\mathcal{B}(\hr)\to\mathcal{B}(\kr): \Phi \textrm{ is linear and } ||\Phi||_{\lozenge}=1  \}=:S_{\lozenge} $. Moreover, it is clear that $\mathcal{C}(\hr,\kr) $ is a compact convex set.\\
A $\tau-$net, $\tau>0$, in $\mathcal{C}(\hr,\kr) $ is a set $\{\mathcal{N}_i  \}_{i=1}^{N}\subset \mathcal{C}(\hr,\kr)$ such that for each $\cn\in \mathcal{C}(\hr,\kr)$ there is at least one $i\in \{1,\cdots,N  \}$ with $||\cn-\cn_i||_{\lozenge}< \tau$. \\
The following basic lemma shows the existence of $\tau-$nets in $\mathcal{C}(\hr,\kr)$ the cardinality of which grows polynomially in $1/\tau$.
\begin{lemma}\label{lemma-tau-nets}
For any $\tau\in (0,1]$ there is a $\tau-$net $\{\cn_i  \}_{i=1}^{N}$ in $\mathcal{C}(\hr,\kr)$ with $N\le (\frac{3}{\tau})^{2(d\cdot d')^2}$, where $d=\dim \hr$ and $d'=\dim \kr$.
\end{lemma}
\emph{Proof.} The lemma is proved by simply imitating the proof of Lemma 2.6 in \cite{milman} where the corresponding result is shown for spheres in arbitrary finite dimensional normed spaces. We give the full argument for convenience.\\
Let $\{ \cm_i \}_{i=1}^{M}$ be an arbitrary subset of $\mathcal{C}(\hr,\kr)$ with the property that
\[||\cm_i-\cm_j||_{\lozenge}\ge\tau,  \]
for all $i\neq j$, $i,j\in\{1,\ldots,M  \}$. We will establish an upper bound on the integer $M$ now.\\
The open balls $B_{\lozenge}(\cm_i,\frac{\tau}{2})$, $i=1,\ldots,M$, with centers at $\cm_i$ and radii $\tau/2$ are mutually disjoint and are contained in the ball $B_{\lozenge}(0,1+\frac{\tau}{2})$ since $\mathcal{C}(\hr,\kr)\subset S_{\lozenge} $. So,
\begin{equation}\label{balls-relation}
 \bigcup_{i=1}^{M}B_{\lozenge}(\cm_i,\frac{\tau}{2})\subset B_{\lozenge}(0,1+\frac{\tau}{2}).
\end{equation}
Let $\mu$ be the Borel-Lebesgue measure (or equivalently the Haar measure) on $(\mathcal{B}(\mathcal{B}(\hr),\mathcal{B}(\kr) ),\Sigma_{Borel})$ where $\mathcal{B}(\mathcal{B}(\hr),\mathcal{B}(\kr) ) $ denotes the set of linear maps from $\mathcal{B}(\hr) $ to $\mathcal{B}(\kr) $ and $\Sigma_{Borel}$ is the $\sigma-$algebra of Borel sets. Computing the volume of the sets in (\ref{balls-relation}) we obtain
\[ M \cdot \left(\frac{\tau}{2}\right)^{2(d\cdot d')^2}\mu(B_{\lozenge}(0,1))\le \left(1+\frac{\tau}{2}\right)^{2(d\cdot d')^2}\mu(B_{\lozenge}(0,1)) ,  \]
where $B_{\lozenge}(0,1) $ is the open unit ball with respect to the $\lozenge-$norm and $2(d \cdot d')^2$ is the dimension of $\mathcal{B}(\mathcal{B}(\hr),\mathcal{B}(\kr) ) $ as a vector space over the field $\rr$. This last inequality is equivalent to
\begin{equation}\label{ball-counting}
  M\le \left(1+\frac{2}{\tau}\right)^{2(d\cdot d')^2}.
\end{equation}
Now, let $\{ \cn_i \}_{i=1}^{N}$ be a \emph{maximal} set in $\mathcal{C}(\hr,\kr) $ with the property that $||\cn_i-\cn_j||_{\lozenge}\ge\tau $ for all $i\neq j$. Then, clearly, $\{ \cn_i \}_{i=1}^{N} $ is a $\tau-$net and (\ref{ball-counting}) holds. Due to our assumption that $\tau\in (0,1]$ we obtain
\[N\le \left(1+\frac{2}{\tau}\right)^{2(d\cdot d')^2} \le \left(\frac{3}{\tau}\right)^{2(d\cdot d')^2} \]
and we are done.
\begin{flushright}$\Box$\end{flushright}
\subsection{\label{subsec:compound-approx}Discrete Approximation of Compound Quantum Channels}
Let $\fri\subseteq \mathcal{C}(\hr,\kr)$ be an arbitrary set. Starting from a $\tau/2-$net $\mathfrak{N}:=\{ \cn_i \}_{i=1}^{N}$ with $N\le (\frac{6}{\tau})^{2(d\cdot d')^2} $ as in Lemma \ref{lemma-tau-nets} we can build a $\tau/2-$net $\fri'_{\tau}$ that is adapted to the set $\fri$ given by
\begin{equation}\label{adapted-net-prel}
  \fri'_{\tau}:=\left\{\cn_i\in \mathfrak{N}: \exists \cn\in \fri \textrm{ with }||\cn-\cn_i||_{\lozenge}<\tau/2  \right\}.
\end{equation}
Let $\mathcal{U}\in\mathcal{C}(\hr,\kr) $ be the useless channel given by $\mathcal{U}(\rho):=\frac{1}{d'}\idn_{\kr}$, $\rho\in \mathcal{S}(\hr)$, and consider
\begin{equation}\label{adapted-net}
  \fri_{\tau}:=\left\{(1-\frac{\tau}{2})\cn+\frac{\tau}{2} \mathcal{U}:\cn\in\fri'_{\tau}  \right\},
\end{equation}
where $\fri'_{\tau}$ is defined in (\ref{adapted-net-prel}).
For $\fri\subseteq \mathcal{C}(\hr,\kr) $ we introduce the abbreviation
\[I_{c}(\rho, \fri ):=\inf_{\cn\in \fri}I_{c}(\rho,\cn),  \]
for $\rho\in \cs (\hr)$.
We list a few more or less obvious results in the following lemma that will be needed in the sequel.
\begin{lemma}\label{lemma-approx-properties}
Let $\fri\subseteq \mathcal{C}(\hr,\kr) $. For each positive $\tau\le \frac{1}{e}$ let $\fri_{\tau}$ be the finite set of channels defined in (\ref{adapted-net}).
\begin{enumerate}
\item $|\fri_{\tau}|\le (\frac{6}{\tau})^{2(d\cdot d')^2} $
\item For $\cn\in \fri$ there is $\cn_i\in \fri_{\tau}$ with
\begin{equation}\label{eq:approx-prop-1}
||\cn^{\otimes l}-\cn_i^{\otimes l}||_{\lozenge}< l\tau.
\end{equation}
Consequently, for $\cn$, $\cn_i$, and any CPTP map $\crr: \mathcal{B}(\kr)^{\otimes l}\to\mathcal{B}(\mathcal{F})$ the relation
\begin{equation}\label{eq:approx-prop-2}
 |F_e(\rho,\crr\circ \cn^{\otimes l})- F_e(\rho,\crr\circ \cn_i^{\otimes l})|<l\tau
\end{equation}
holds for all $\rho\in\cs(\hr^{\otimes l})$ and $l\in \nn$.
\item For all $\rho\in\cs (\hr)$ we have
  \begin{equation}\label{eq:approx-prop-3}
    |I_{c}(\rho,\fri)-I_{c}(\rho,\fri_{\tau})|\le \tau+2\tau\log \frac{d}{\tau}.
  \end{equation}

\end{enumerate}
\end{lemma}
\emph{Proof.} 1. This is clear from definition of $\fri_\tau$.\\
2. It is clear from construction of $\fri_{\tau}$ that there is at least one $\cn_i\in \fri_{\tau}$ with
\[||\cn-\cn_i||_{\lozenge}< \tau.  \]
We know from \cite{kitaev, paulsen} that $||\crr_1\otimes \crr_2||_{\lozenge}=||\crr_1||_{\lozenge}\cdot||\crr_2||_{\lozenge}$ holds, i.e. $\lozenge-$norm is multiplicative. The inequality (\ref{eq:approx-prop-1}) is easily seen using repeatedly the tensor identity
\[a_1\otimes b_1-a_2\otimes b_2=a_1\otimes (b_1-b_2)+(a_1-a_2)\otimes b_2,  \]
 the multiplicativity of the $\lozenge-$norm, and the fact that $||\crr||_{\lozenge}=1$ for all CPTP maps.  \\
Let $\psi\in \hr^{\otimes l}\otimes \hr^{\otimes l}$ be a purification of $\rho\in\cs(\hr^{\otimes l}) $. Let us denote the left hand side of (\ref{eq:approx-prop-2}) by $\triangle F_e $. By the definition of the entanglement fidelity we have
\[\triangle F_e=|\langle \psi, id_{d}^{\otimes l}\otimes (\crr\circ (\cn^{\otimes l}-\cn_i^{\otimes l}))(|\psi\rangle\langle \psi| )\psi\rangle    |.  \]
 An application of the Cauchy-Schwarz inequality shows that
\begin{eqnarray*}
  \triangle F_e&\le & || id_d^{\otimes l}\otimes (\crr\circ (\cn^{\otimes l}-\cn_i^{\otimes l}))(|\psi\rangle\langle \psi| )\psi ||\\
&\le& || id_d^{\otimes l}\otimes (\crr\circ (\cn^{\otimes l}-\cn_i^{\otimes l}))(|\psi\rangle\langle \psi| )||_{\infty}\\
&\le& ||id_d^{\otimes l}\otimes (\crr\circ (\cn^{\otimes l}-\cn_i^{\otimes l}))(|\psi\rangle\langle \psi| )||_1\\
&=& ||(id_d^{\otimes l}\otimes \crr)\circ(id_d\otimes  (\cn^{\otimes l}-\cn_i^{\otimes l})))(|\psi\rangle\langle \psi| )||_1\\
&\le& ||\crr||_{\lozenge}||\cn^{\otimes l}-\cn_i^{\otimes l}||_{\lozenge}|||\psi\rangle\langle \psi|||_1\\
&<& l\tau,
\end{eqnarray*}
where we have used  $||\crr||_{\lozenge}=1 $, $|||\psi\rangle\langle \psi|||_1=1 $, and (\ref{eq:approx-prop-1}).\\
3. The proof of (\ref{eq:approx-prop-3}) is based on Fannes inequality \cite{fannes} and uses merely standard conclusions. So, we will confine ourselves to a brief outline of the argument. Fannes inequality states that $ |S(\sigma_1)-S(\sigma_2)|\le \tau\log d-\tau\log\tau$ for all density operators with $||\sigma_1-\sigma_2||_1\le \tau\le 1/e$. To the given $\tau$ we can always find an $\cn'\in \fri$ with
\begin{equation}\label{eq:approx-1}
I_c(\rho,\cn')\le I_{c}(\rho,\fri)+\tau .
\end{equation}
On the other hand there is $\cn_i\in\fri_{\tau}$ with $||\cn'-\cn_i||_{\lozenge}<\tau$. This implies immediately
 \[||\cn'(\rho)-\cn_i(\rho)||_1< \tau , \]
and
\[||id_{d}\otimes \cn' (|\psi\rangle\langle \psi|)-  id_{d}\otimes \cn_i (|\psi\rangle\langle \psi|)||_1< \tau\]
by the definition of  $\lozenge-$norm where $\psi\in \hr\otimes \hr$ is a purification of $\rho\in\cs (\hr)$. Since
\[ I_{c}(\rho,\cn')=S(\cn'(\rho))-S(id_{d}\otimes \cn' (|\psi\rangle\langle \psi|) ) \]
with a similar relation for $\cn_i$, an application of Fannes inequality leads to
\[ I_{c}(\rho,\cn_i)\le I_{c}(\rho,\cn')+2(\tau\log d-\tau\log\tau). \]
This and (\ref{eq:approx-1}) show that
\[I_{c}(\rho,\fri_{\tau})\le I_{c}(\rho,\fri)+\tau +2(\tau\log d-\tau\log\tau). \]
The inequality
\[I_{c}(\rho,\fri)\le I_{c}(\rho,\fri_{\tau})+2(\tau\log d-\tau\log\tau)   \]
is shown in a similar vein.
\begin{flushright}$\Box$\end{flushright}

\section{\label{sec:direct-part}Direct Part of the Coding Theorem: General Case with Informed Decoder }
In this final section we will utilize our coding results for finite $\fri$ and the discretization techniques from the previous section to establish the coding theorem for arbitrary $\fri$ with informed decoder.
\begin{theorem}\label{direct-coding-general}
Let $\fri\in \mathcal{C}(\hr,\kr)$ be an arbitrary compound channel and let $\pi_{\gr}$ be the maximally mixed state associated with a subspace $\gr\subset \hr$. Then
\[Q_{ID}(\fri)\ge \inf_{\cn\in \fri}I_c(\pi_{\gr},\cn).  \]
\end{theorem}
\emph{Proof.} We consider two subspaces $\fr_l, \gr^{\otimes l}$ of $\hr^{\otimes l}$ with $\fr_l\subset\gr^{\otimes l}\subset \hr^{\otimes l}$. Let $k_l:=\dim \fr_l$ and we denote as before the associated maximally mixed states on $\fr_l$ and $\gr$ by $\pi_{\fr_l}$ and  $\pi_{\gr}$.\\
If $\inf_{\cn\in \fri}I_c(\pi_{\gr},\cn)\le 0 $ there is nothing to prove. Therefore we suppose that
\[ \inf_{\cn\in \fri}I_c(\pi_{\gr},\cn)>0 \]
holds. We will show that for each $\eps\in (0, \inf_{\cn\in \fri}I_c(\pi_{\gr},\cn) )$ the number
\[\inf_{\cn\in \fri}I_c(\pi_{\gr},\cn)-\eps  \]
is an achievable rate.\\
Let $\tau>0$ with $\tau\le \frac{1}{e}$ which will be specified later. We consider the finite set of channels $\fri_{\tau}:=\{\cn_1,\ldots, \cn_{N_{\tau}}\}$ associated to $\fri$ given in (\ref{adapted-net}) with the properties listed in lemma \ref{lemma-approx-properties}. We can conclude from the proof of theorem \ref{lemma-direct-finite-1} that for each $l\in \nn$ there is a subspace $\fr_l\subset \gr^{\otimes l}$ of dimension
\begin{equation}
k_l= \lfloor  2^{l(\min_{i\in \{1,\ldots, N_{\tau}  \}}I_c(\pi_{\gr},\cn_i)-\frac{\eps}{2}  )}  \rfloor
\end{equation}
and for each $\cn_i\in \fri_\tau$ a recovery operation $\mathcal{R}_i$ such that
\begin{eqnarray}\label{general-direct-2}
F_e(\pi_{\fr_l}, \mathcal{R}_i\circ \cn_i^{\otimes l} )&\ge& 1-N_{\tau}(2^{-l(c\delta^2-h(l))}+2^{-l(c'\delta^2-h'(l))}\nonumber\\
                                           & & +2N_\tau\sqrt{2^{-l(\frac{\eps}{2}-\gamma(\delta)-3\vphi(\delta))}}),
\end{eqnarray}
where we have chosen $\tau>0$ and $\delta>0$ small enough to ensure that
\[ \min_{i\in \{1,\ldots, N_{\tau}  \}}I_c(\pi_{\gr},\cn_i)-\frac{\eps}{2}>0,  \]
and
\[ \frac{\eps}{2}-\gamma(\delta)-3\vphi(\delta)>0. \]
By our construction of $\fri_{\tau}$ we can find to each $\cn\in \fri$ at least one $\cn_i\in \fri_{\tau}$ with
\begin{equation}\label{general-direct-3}
 |F_{e}(\pi_{\fr_l},\mathcal{R}_i\circ \cn_i^{\otimes l} )- F_{e}(\pi_{\fr_l},\mathcal{R}_i\circ \cn^{\otimes l} )     | \le l\cdot \tau
\end{equation}
according to lemma \ref{lemma-approx-properties}. Note that by the last claim of lemma \ref{lemma-approx-properties} we also have the following estimate on the dimension $k_l$ of the subspace $\fr_l$:
\begin{equation}\label{general-direct-4}
k_l \ge \lfloor  2^{l (\inf_{\cn\in \fri}I_c(\pi_{\gr},\cn )-\frac{\eps}{2}-\tau -2\tau\log\frac{d}{\tau}}   \rfloor.
\end{equation}
The relations (\ref{general-direct-3}), (\ref{general-direct-2}), and (\ref{general-direct-4}) lead to the following conclusion:  For each $l\in \nn$ we can find a subspace $\fr_l\subset \gr^{\otimes l}\subset\hr^{\otimes l}$ of dimension
\begin{equation}\label{general-direct-5}
k_l \ge \lfloor  2^{l (\inf_{\cn\in \fri}I_c(\pi_{\gr},\cn )-\frac{\eps}{2}-\tau -2\tau\log\frac{d}{\tau}}   \rfloor
\end{equation}
such that for each $\cn\in \fri$ there is a recovery operation $\mathcal{R}_{\cn}$ with
\begin{eqnarray}\label{general-direct-6}
F_e(\pi_{\fr_l}, \mathcal{R}_{\cn}\circ \cn^{\otimes l} )&\ge& 1-N_{\tau}(2N_\tau\sqrt{2^{-l(\frac{\eps}{2}-\gamma(\delta)-3\vphi(\delta))}}\nonumber\\
                                           &&+2^{-l(c\delta^2-h(l))}+2^{-l(c'\delta^2-h'(l))})\nonumber\\
                                           &&-l \tau.
\end{eqnarray}
Finally if we choose our approximation parameter $\tau$ in dependence of $l\in\nn$ we can ensure that the entanglement fidelity of our code approaches one: Taking e.g. any sequence $(\tau_l)_{l\in \nn}$ with $\tau_l\searrow 0$ as $l$ tends to $\infty$, and $\lim_{l\to\infty}l\cdot \tau_l=0$ we see immediately that the RHS of (\ref{general-direct-6}) tends to one, and that the RHS of (\ref{general-direct-5}) can be lower-bounded by
\[ \lfloor2^{l(\inf_{\cn\in \fri}I_c(\pi_{\gr},\cn )-\eps  )}\rfloor  \]
for all sufficiently large $l\in\nn$.
\begin{flushright}$\Box$\end{flushright}
The following lemma is the compound analog of a result discovered by Bennett, Shor, Smolin, and Thapliyal in \cite{bsst} (BSST lemma for short). We will use the following abbreviations: For any $\fri\subset \mathcal{C}(\hr,\kr)$ and $l\in \nn$ we set
\[\fri^{\otimes l}:=\left\{\cn^{\otimes l}: \cn\in \fri  \right\}. \]
Recall also our earlier shortcut notation
\[I_c(\rho,\fri)=\inf_{\cn\in \fri}I_c(\rho,\cn)  \]
for $\rho\in \mathcal{S}(\hr)$.
\begin{lemma}[Compound BSST Lemma]\label{compound-bsst-lemma}
Let $\fri\subset\mathcal{C}(\hr,\kr) $ be an arbitrary set of channels. For any $\rho\in \mathcal{S}(\hr)$ let $q_{\delta,l}\in\mathcal{B}(\hr^{\otimes l})$ be the frequency-typical projection of $\rho$ and set
\[ \pi_{\delta,l}:=\frac{q_{\delta,l}}{\textrm{tr}(q_{\delta,l})}\in\mathcal{S}(\hr^{\otimes l}). \]
Then there is a positive sequence $(\delta_l)_{l\in\nn}$ satisfying $\lim_{l\to\infty}\delta_l=0$ with
\[ \lim_{l\to\infty}\frac{1}{l}\inf_{\cn\in \fri}I_c(\pi_{\delta_l,l},\cn^{\otimes l} )=\inf_{\cn\in\fri}I_c(\rho,\cn).  \]

\end{lemma}
\emph{Proof.} The proof is via reduction to Holevo's proof \cite{holevo-ent-ass-cap} of the BSST lemma for single channel supplemented by a discretization argument. As in the proof of theorem \ref{direct-coding-general} we choose a decreasing sequence $(\tau_l)_{l\in\nn}$, with $\tau_l>0$, $\lim_{l\to\infty}l\tau_l=0$, and consider the finite set of channels $\fri_{\tau_l}=\{\cn_1,\ldots,\cn_{N_{\tau_l}}  \}$ defined in (\ref{adapted-net}) associated to $\fri$.\\
By our construction of the set $\fri_{\tau_l}$ we know that
\begin{equation}\label{compound-bsst-1}
  \cn_{i}(\rho)\ge \frac{\tau_l}{d'2}\idn_{\kr}
\end{equation}
holds for all $i\in \{1,\ldots, N_{\tau_l}\}$, which implies
\begin{equation}\label{compound-bsst-2}
  \log \cn_{i}(\rho)\ge \log \left(\frac{\tau_l}{d'2}\right)\idn_{\kr}
\end{equation}
uniformly in $i\in \{1,\ldots, N_{\tau_l}\} $. On the other hand let $\fri_{\tau_l,e}=\{\mathcal{E}_1,\ldots \mathcal{E}_{N_{\tau_l}}  \}\subset\mathcal{C}(\hr,\hr_e)$ denote the complementary set of channels associated to $\fri_{\tau_l}$. We alter $\fri_{\tau_l,e}$ by mixing a part of useless channel $\mathcal{U}_e\in\mathcal{C}(\hr,\hr_e)$ to each $\mathcal{E}_i\in\fri_{\tau_l,e} $, i.e. we set
\[\fri'_{\tau_l,e}:=\left(1-\frac{\tau_l}{2}\right)\fri_{\tau_l,e}+\frac{\tau_l}{2}\mathcal{U}_e . \]
Note that then for each $\mathcal{E}'_i\in \fri'_{\tau_l,e}$
\[ \mathcal{E}'_i(\rho)\ge \frac{\tau_l}{2\dim (\hr_e)}\idn_{\hr_e},  \]
and consequently
\begin{equation}\label{compound-bsst-2a}
  \log \mathcal{E}'_i(\rho)\ge \log \left(\frac{\tau_l}{2\dim (\hr_e)}  \right)\idn_{\hr_e}.
\end{equation}

Applying Holevo's argument from \cite{holevo-ent-ass-cap} to each channel from $\fri_{\tau_l}$ and $\fri'_{\tau_l,e} $ with our notation from lemma \ref{lemma-typical-1} and uniform bounds (\ref{compound-bsst-2}), (\ref{compound-bsst-2a}) we obtain for each $i\in \{1,\ldots, N_{\tau_l}\} $
\begin{eqnarray}\label{compound-bsst-3}
  \left|\frac{1}{l}S(\mathcal{T}_i^{\otimes l}(\pi_{\delta,l} ) )-S(\mathcal{T}_i(\rho))  \right|&\le& -\frac{1}{l}\log \eta_l(\delta)+2\vphi(\delta)\nonumber\\
&& - d\delta \log \left( \frac{\tau_l}{2 D}  \right) \nonumber\\
&=:& \Theta_l(\delta,D)
\end{eqnarray}
where $(\mathcal{T}_i,D)\in\{(\cn_i,d'),(\mathcal{E}'_i,\dim \hr_e)  \}$.\\
Since for each $i\in\{1,\ldots, N_{\tau_l}\} $
\[||\mathcal{E}_i-\mathcal{E}'_i||_{\lozenge}\le \tau_l,  \]
we obtain
\[ ||\mathcal{E}_i^{\otimes l}-{\mathcal{E}'_{i}}^{\otimes l}||_{\lozenge}\le l\tau_l. \]
Hence choosing $l$ sufficiently large we can ensure that $l\tau_l\le \frac{1}{e}$ and an application of Fannes inequality shows that
\begin{equation}\label{compound-bsst-4a}
  |S(\mathcal{E}'_i(\rho))-S(\mathcal{E}_i(\rho) )  |\le \tau_l\log\frac{\dim \hr_e}{\tau_l},
\end{equation}
and
\begin{equation}\label{compound-bsst-4b}
\left| \frac{1}{l}S\left({\mathcal{E}'_i}^{\otimes l}(\pi_{\delta,l})  \right) -  \frac{1}{l}S\left(\mathcal{E}_i^{\otimes l}(\pi_{\delta,l})  \right) \right| \le l\tau_l \log \frac{\dim \hr_e}{l\tau_l}.
\end{equation}
Inequalities (\ref{compound-bsst-3}), (\ref{compound-bsst-4a}), and (\ref{compound-bsst-4b}) show that
\begin{eqnarray}\label{compound-bsst-5}
 \left|\frac{1}{l}I_c(\pi_{\delta,l}, \cn_i^{\otimes l}  )-I_c(\rho,\cn_i)         \right|&\le& \Theta_l(\delta,d')+\Theta_l(\delta, \dim \hr_e)\nonumber\\
&&+ \tau_l\log\frac{\dim \hr_e}{\tau_l}\nonumber\\
&&+l\tau_l \log \frac{\dim \hr_e}{l\tau_l}\nonumber\\
&=:& \Delta_l(\delta,d',\dim \hr_e)
\end{eqnarray}
for each $i\in \{1,\ldots, N_{\tau_l}\} $. It is then easily seen utilizing (\ref{compound-bsst-5}) that
\begin{equation}\label{compound-bsst-6}
  \left|\frac{1}{l}I_c(\pi_{\delta,l}, \fri_{\tau_l}^{\otimes l}  )-I_c(\rho,\fri_{\tau_l})    \right|\le \Delta_l(\delta,d',\dim \hr_e).
\end{equation}
Applying (\ref{eq:approx-prop-3} ) to $\rho$, $\fri$ and $\pi_{\delta,l}$, $\fri^{\otimes l}$ we obtain
\begin{equation}\label{compound-bsst-7}
  \left|I_c(\rho,\fri_{\tau_l})-I_c(\rho,\fri)     \right|\le \tau_l+2\tau_l\log \frac{d}{\tau_l},
\end{equation}
and
\begin{equation}\label{compound-bsst-8}
  \left| \frac{1}{l}I_c(\pi_{\delta,l},\fri_{\tau_l}^{\otimes l})- \frac{1}{l}I_c(\pi_{\delta,l},\fri^{\otimes l})  \right|\le \tau_l+2l\tau_l \log \frac{d}{l\tau_l}.
\end{equation}
Using triangle inequality, (\ref{compound-bsst-6}), (\ref{compound-bsst-7}), and (\ref{compound-bsst-8})  we see that
\begin{eqnarray}\label{compound-bsst-9}
  \left| \frac{1}{l}I_c(\pi_{\delta,l},\fri^{\otimes l})-  I_c(\rho,\fri)   \right|&\le& \Delta_l(\delta,d',\dim \hr_e)\nonumber\\
&& + \tau_l+2l\tau_l \log \frac{d}{l\tau_l}\nonumber\\
&&+ \tau_l+2\tau_l\log \frac{d}{\tau_l}.
\end{eqnarray}
We are done now, since for any positive sequence $(\delta_l)_{l\in\nn}$ with $\lim_{l\to\infty}\delta_l=0$, $\lim_{l\to\infty}\eta_l(\delta_l)=1$, and
$\lim_{l\to\infty} \delta_l\log \tau_l=0$ we can conclude from (\ref{compound-bsst-9}) that
\[ \lim_{l\to\infty}\frac{1}{l}\inf_{\cn\in \fri}I_c(\pi_{\delta_l,l},\cn^{\otimes l} )=\inf_{\cn\in\fri}I_c(\rho,\cn).  \]
holds.
\begin{flushright}$\Box$\end{flushright}
It is easy now to prove the direct part of the coding theorem.\\
First observe that for each $k\in\nn$
\begin{equation}\label{direct-coding-inf-dec-1}
Q_{ID}(\fri^{\otimes k})=k Q_{ID}(\fri)
\end{equation}
holds. For any fixed $\rho\in\mathcal{S}(\hr^{\otimes m})$ let $q_{\delta,l}\in \mathcal{B}(\hr^{\otimes ml})$ be the frequency-typical projection of $\rho$ and set $\pi_{\delta,l}=\frac{q_{\delta,l}}{\textrm{tr}(q_{\delta,l})}$. Then according to theorem \ref{direct-coding-general} for each $\delta \in (0,1/2)$ we have
\begin{equation}\label{direct-coding-inf-dec-2}
Q_{ID}(\fri^{\otimes ml})\ge I_c(\pi_{\delta,l},\fri^{\otimes ml}).
\end{equation}
From (\ref{direct-coding-inf-dec-1}), (\ref{direct-coding-inf-dec-2}), and lemma \ref{compound-bsst-lemma} we obtain
\begin{eqnarray}\label{direct-coding-inf-dec-3}
 Q_{ID}(\fri)&=& \frac{1}{m}\lim_{l\to\infty}\frac{1}{l}Q_{ID}(\fri^{\otimes ml})\nonumber\\
        &\ge& \frac{1}{m}  \lim_{l\to\infty}\frac{1}{l}\inf_{\cn\in \fri}I_c(\pi_{\delta_l,l},(\cn^{\otimes m})^{\otimes l} )\nonumber\\
        &=&  \frac{1}{m}I_c(\rho, \fri^{\otimes m}).
\end{eqnarray}
Obviously, (\ref{direct-coding-inf-dec-3}) implies the following theorem:
\begin{theorem}[Direct Part: Informed Decoder]\label{direct-coding-inf-dec}
Let $\fri\subset\mathcal{C}(\hr,\kr)$. Then the capacity of the compound quantum channel built up from $\fri$ in the scenario with informed decoder satisfies
\[Q_{ID}(\fri)\ge \lim_{l\to\infty}\frac{1}{l} \max_{\rho\in\mathcal{S}(\hr^{\otimes l})}\inf_{\cn\in\fri}I_c(\rho,\cn^{\otimes l}).  \]

\end{theorem}
Since the converse to the coding theorem \ref{direct-coding-inf-dec} follows easily from the converse part to the coding theorem for \emph{single memoryless channel} \cite{bkn} we can state our capacity result as follows:
\begin{theorem}[Capacity with Informed Decoder]\label{coding-theorem-inf-dec}
The capacity of the compound channel built up from a set $\fri\subset\mathcal{C}(\hr,\kr) $ with informed decoder is given by
\[ Q_{ID}(\fri)= \lim_{l\to\infty}\frac{1}{l} \max_{\rho\in\mathcal{S}(\hr^{\otimes l})}\inf_{\cn\in\fri}I_c(\rho,\cn^{\otimes l}). \]
\end{theorem}


\begin{acknowledgments}
IB would like to thank Nilanjana Datta for explaining to him parts of her joint work with Dorlas \cite{datta-dorlas} and a very stimulating correspondence on classical capacities of compound quantum channels.\\
IB is supported by the Deutsche Forschungsgemeinschaft (DFG) via project ``Entropie und Kodierung gro\ss er Quanten-Informationssysteme'' at the TU Berlin. HB and JN are grateful for the support by the TU Berlin through the fund for basic research.
\end{acknowledgments}


\end{document}